\documentclass[12pt]{article}
\usepackage[left=2cm, right=2cm, bottom=2cm ,top = 2cm, footskip=1cm]{geometry}
\usepackage{setspace}
\usepackage{chngcntr}

\usepackage[authoryear,sort&compress]{natbib}
\setcitestyle{open={(},close={)}}

\usepackage{graphicx}
\usepackage{amsmath}
\usepackage[section]{placeins}

\usepackage[nolist]{acronym} 

\title{Seizure pathways and seizure durations can vary independently within individual patients with focal epilepsy}
\author{Gabrielle M. Schroeder$^{1}$, Fahmida A. Chowdhury$^{3}$, Mark J. Cook$^{4}$, Beate Diehl$^{3}$,\\ John S. Duncan$^{3}$, Philippa J. Karoly$^{4,5}$, Peter N. Taylor$^{1,2,3}$, Yujiang Wang$^{1,2,3*}$}

\begin{document}
\maketitle

\begin{enumerate}
\item{CNNP Lab (www.cnnp-lab.com), Interdisciplinary Computing and Complex BioSystems Group, School of Computing, Newcastle University, Newcastle upon Tyne, United Kingdom}
\item{Faculty of Medical Sciences, Newcastle University, Newcastle upon Tyne, United Kingdom}
\item{UCL Queen Square Institute of Neurology, Queen Square, London, United Kingdom}
\item{Graeme Clark Institute and St Vincent's Hospital, University of Melbourne, Melbourne, Victoria, Australia}
\item{Department of Biomedical Engineering, University of Melbourne, Melbourne, Victoria, Australia}
\end{enumerate}

\begin{center}
* Yujiang.Wang@newcastle.ac.uk    
\end{center}

\newpage

\begin{acronym}

\acro{ARI}{adjusted Rand index}
\acro{AED}{antiepileptic drug}
\acro{DTW}{dynamic time warping} 
\acro{EMU}{epilepsy monitoring unit}
\acro{FDR}{false discovery rate}
\acro{EEG}{electroencephalographic}
\acro{iEEG}{intracranial EEG}
\acro{ILAE}{International League Against Epilepsy}
\acro{MDS}{multidimensional scaling}
\acro{NMF}{non-negative matrix factorisation}
\acro{UCLH}{University College London Hospital}
\acro{UPGMA}{unweighted pair group method with arithmetic mean}

\end{acronym}

\section{Abstract}

A seizure's electrographic dynamics are characterised by its spatiotemporal evolution, also termed dynamical ``pathway'', and the time it takes to complete that pathway, which results in the seizure’s duration. Both seizure pathways and durations can vary within the same patient, producing seizures with different dynamics, severity, and clinical implications. However, it is unclear whether seizures following the same pathway will have the same duration or if these features can vary independently. We compared within-subject variability in these seizure features using 1) epilepsy monitoring unit intracranial EEG (iEEG) recordings of 31 patients (mean 6.7 days, 16.5 seizures/subject), 2) NeuroVista chronic iEEG recordings of 10 patients (mean 521.2 days, 252.6 seizures/subject), and 3) chronic iEEG recordings of 3 dogs with focal-onset seizures (mean 324.4 days, 62.3 seizures/subject). While the strength of the relationship between seizure pathways and durations was highly subject-specific, in most subjects, changes in seizure pathways were only weakly to moderately associated with differences in seizure durations. The relationship between seizure pathways and durations was weakened by seizures that 1) had a common pathway, but different durations (``elastic pathways”), or 2) had similar durations, but followed different pathways (``duplicate durations”). Even in subjects with distinct populations of short and long seizures, seizure durations were not a reliable indicator of different seizure pathways. These findings suggest that seizure pathways and durations are modulated by different processes. Uncovering such modulators may reveal novel therapeutic targets for reducing seizure duration and severity. 


\section{Introduction}

Many health conditions are challenging to treat due to changes in symptoms and disease severity over time \citep{Fava2003,Frey2011,Barends2020}. Recent research has emphasised that temporal fluctuations are also an important consideration in focal epilepsy, as seizures can change over time within the same patient \citep{Schroeder2020,Saggio2020,Cook2016,Karoly2018b,Salami2020,Salami2021}. Specifically, a patient's spatiotemporal seizure dynamics can vary in two main ways. 
First, the evolution of pathological activity, as measured by \ac{EEG} recordings, can differ from seizure to seizure. These seizure evolutions can be described mathematically with various computational models, including functional networks \citep{Burns2014,Schroeder2020} and neural mass models \citep{Karoly2018b}, that capture specific dynamical seizure properties. Using such approaches, each seizure evolution can be conceptualised as a pathway through the chosen feature space \citep{Nevado-Holgado2012,Karoly2018b,Schroeder2020,Wendling2002,Jirsa2014}. Second, each seizure is also characterised by its duration, which is commonly defined as the amount of time that elapses from its electrographic start to finish \citep{Halford2015,Kaufmann2020,Dobesberger2015,Kim2011dur}. Together, these features describe both the sequence of brain activity during a seizure as well as the amount of time that it takes to complete that sequence. 

Both seizure pathways and seizure durations are related to seizure clinical symptoms and severity \citep{Kaufmann2020,Dobesberger2015,Kim2011dur,Schroeder2020,Cramer2001}. Certain types of diversity in seizure pathways, such as multifocal onsets \citep{Spencer1981,Rossi1994,Schmeiser2017} and variable recruitment patterns \citep{Martinet2015,Farooque2014}, are associated with worse outcomes following epilepsy surgery. Additionally, seizures can be more difficult to predict in patients with distinct populations of short and long seizures \citep{Cook2016}. However, despite the clinical relevance of seizure pathways and durations, little is known about how these features interact. In particular, it is unclear whether variability in seizure duration arises purely from changes in seizure pathways, or whether pathways and durations can vary independently within the same patient. 

Some previous studies suggest that seizure pathways and durations are linked, with different durations corresponding to seizures with distinct pathways. First, seizure duration often differs between different \ac{ILAE} clinical seizure types, which classify seizures based on clinical symptoms and \ac{EEG} features \citep{Fisher2017} and are also associated with changes in functional networks \citep{Schroeder2020,Schindler2007a,Burns2014}.
Seizure types with more severe clinical manifestations have also been reported to last longer; for example, focal seizures that progress to bilateral tonic-clonic seizures tend to have longer durations than seizures that remain focal \citep{Kaufmann2020,Dobesberger2015}, and focal seizures tend to be longer if they involve loss of awareness \citep{Kim2011dur,Dobesberger2015}. Meanwhile, analysis of chronic \ac{iEEG} recordings suggests that seizures with different durations have similar onsets, but different terminations \citep{Karoly2018b}. Additionally, there is evidence that distinct populations of short and long seizures correspond to different seizure pathways with characteristic durations \citep{Cook2016, Karoly2018b}. These findings suggest that seizure pathways and durations may co-vary within patients, with different seizure durations serving as a proxy for different seizure pathways.

However, it is also possible that seizure duration is modulated independently of the seizure's pathway. Two seizures could potentially follow the same pathway, but have different durations due to variable rates of progression (e.g. by ``dwelling'' in particular EEG activity patterns). In a rodent model, \cite{Wenzel2017} found seizures with consistent recruitment patterns and different rates of seizure spread at a neuronal level, a characteristic termed ``elasticity." To our knowledge, no studies have quantitatively explored such temporal flexibility in seizure pathways in human patients. Nonetheless, within-patient seizures with consistent firing patterns, but small changes in duration, have been observed \citep{Truccolo2011}, suggesting that elasticity in the same seizure pathway may also occur in humans. This mechanism could potentially lead to variable durations among seizures with the same pathway. 

The relationship between seizure pathways and durations has been difficult to investigate due to the lack of an objective measure for comparing seizure pathways. We addressed this need by proposing an approach for quantitatively comparing within-patient seizure pathways, which we used to investigate variability in seizure functional network evolutions \citep{Schroeder2020}. In the present study, we used the same approach to explore if variability in seizure pathways is linked to variability in seizure durations. Our comparison of seizure pathways allowed us to recognise similar pathways even if they progressed at different rates. Thus, we could determine if two seizures shared the same pathway, even if their durations differed. We also extended our analysis to include long-term recordings from NeuroVista patients \citep{Cook2013} and dogs \citep{Howbert2014,Davis2011}, allowing us to analyse the relationship between pathways and durations in subjects with a higher number of recorded seizures that occurred over longer timescales. Our analysis revealed that seizure pathways and durations are not tightly linked in most patients, allowing there to be independent variability in each feature.

\section{Results}

We analysed a total of 3,224 seizures, recorded using \ac{iEEG} (Fig. \ref{F:measures}A), from 
\begin{itemize}
    \item \ac{EMU} patients: 31 patients with focal epilepsy who underwent presurgical monitoring in \ac{EMU}s (average 16.5 seizures/patient)
    \item NeuroVista patients: 10 patients with focal epilepsy who underwent chronic recordings as part of the NeuroVista seizure prediction study \citep{Cook2013} (average 252.6 seizures/patient)
    \item Dogs: 3 dogs with naturally occurring canine epilepsy and focal-onset seizures \citep{Howbert2014,Davis2011} (average 62.3 seizures/subject). 
\end{itemize}

Fig. \ref{F:measures}B shows the \ac{iEEG} recordings of four seizures from an example subject, \ac{EMU} 821. 

\FloatBarrier
\subsection{Quantifying within-subject variability in seizure pathways and seizure durations}

\begin{figure}[hbp]
	\centering
		\includegraphics[width=0.9\textwidth]{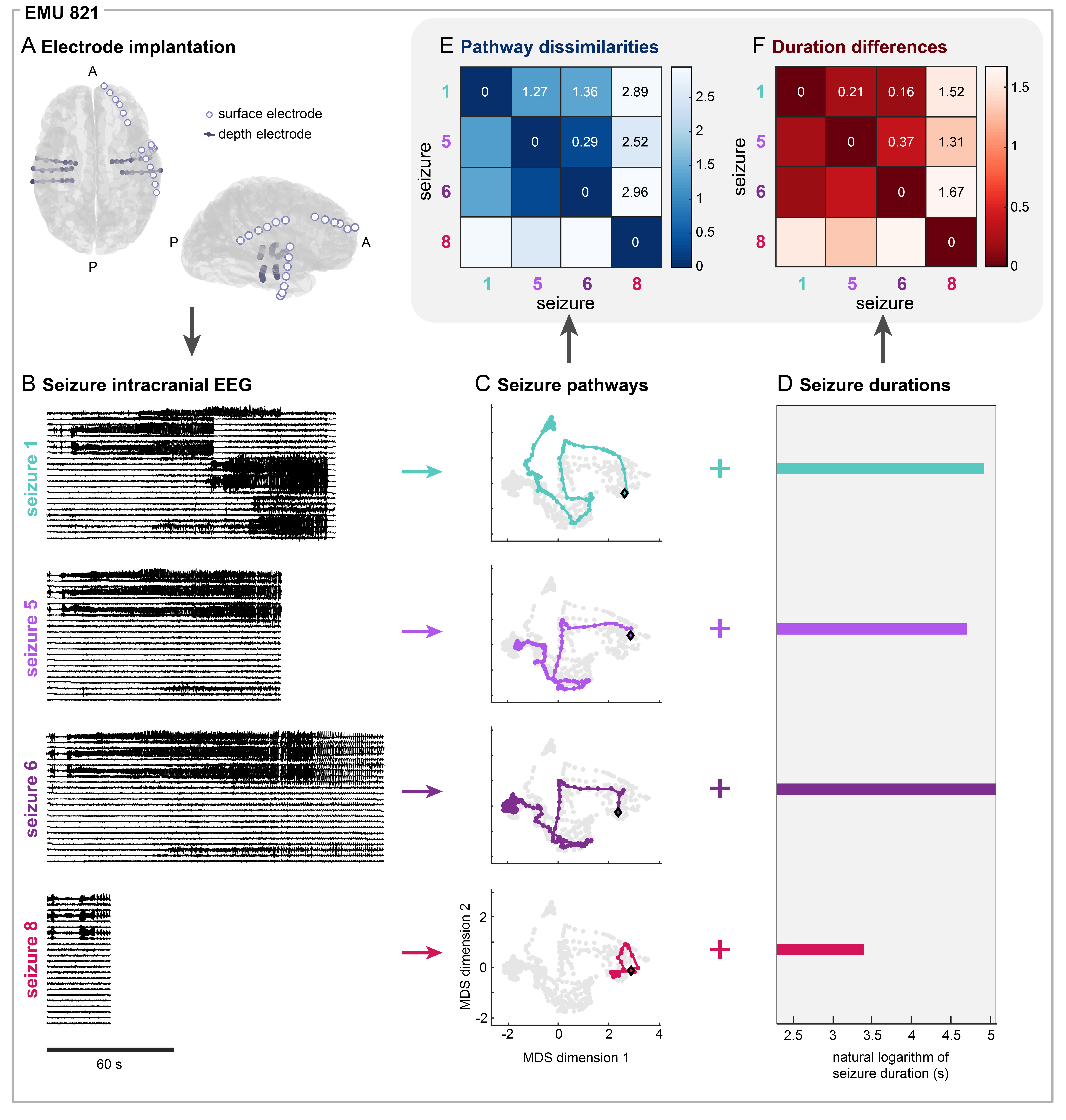}
		\caption[Quantitatively comparing seizure pathways and durations within individual subjects]{\textbf{Quantitatively comparing seizure pathways and durations within individual subjects.} A) Schematic of the electrode implantation for \ac{iEEG} recording of an example subject, \ac{EMU} 821. A = anterior of brain, P = posterior of brain. B) Intracranial EEG of four of \ac{EMU} 821's seizures. The recordings from a representative subset of electrodes are shown. Seizure numbers refer to their chronological order. C) Multidimensional scaling (MDS) embeddings of the corresponding pathways of the example seizures. Each point corresponds to the functional network configuration of a seizure time window, and time windows with more similar network configurations are located closer together in the embedding. Coloured points correspond to time windows that occurred during the example seizure, with the first time window marked with a black diamond and successive time window connected with the coloured line to form the seizure pathway. Time windows that occurred during other seizures are shown in grey for reference. D) The durations of each of the example seizures, shown on a natural logarithm scale. Seizure dynamics were characterised by seizure pathways (C) and seizure durations (D). E) Pairwise pathway dissimilarities and F) duration differences of the example seizures. Both matrices are symmetric.}
		\label{F:measures}
\end{figure}

We described the dynamics of each seizure using two features: 
\begin{enumerate}
	\item The seizure's functional network evolution, which can be considered a \textit{pathway} through the space of possible functional network interactions (Fig. \ref{F:measures}C).
	\item The time it takes the seizure to follow its pathway; i.e., the seizure's \textit{duration} (Fig. \ref{F:measures}D).
\end{enumerate}
For clarity, we will only use the terms \textit{short}/\textit{long} to describe seizure temporal duration and \textit{small}/\textit{large} to describe relative amounts of spatial distances followed by seizure pathways through the functional network space. 

Following our previous work \citep{Schroeder2020}, we described seizure pathways through network space by computing the time-varying (sliding window) coherence between pairs of \ac{iEEG} channels across six frequency bands: delta (1-4 Hz), theta (4-8 Hz), alpha (8-13 Hz), beta (13-30 Hz), gamma (30-80 Hz), and high gamma (80-150 Hz). 

To quantify similarities and differences in seizure pathways, we used \ac{DTW} \citep{Sakoe1978} to compute pairwise dissimilarities between seizures, resulting in a symmetric ``seizure dissimilarity" matrix for each subject \citep{Schroeder2020} (Fig. \ref{F:measures}E). In our case, \ac{DTW} minimised the overall distance between a pair of seizure pathways by selectively stretching parts of each pathway such that similar network configurations were temporally aligned. Therefore, \ac{DTW} allowed us to recognise similar seizure pathways even if the seizures had different durations. We defined the ``pathway dissimilarity" between a pair of seizures as the average distance between their functional connectivity time series after \ac{DTW}. Additionally, to visualise seizure pathways through network space, we used \ac{MDS} to project each subject's seizure network evolutions into a two-dimensional space (Fig. \ref{F:measures}C, also see Methods). Each point in the projection corresponds to a network configuration that occurred during the seizure, and time points with more similar network configurations tend to be placed closer together. 

To compare variability in seizure pathways to variability in seizure durations, we quantified the pairwise differences in each subject's seizure durations. As in previous work \citep{Cook2016}, we first computed the natural logarithm of each seizure duration (Fig. \ref{F:measures}D). We then computed the pairwise absolute differences between the transformed seizure durations, resulting in a symmetric ``duration difference" matrix for each subject (Fig. \ref{F:measures}F). Due to the properties of logarithms, our measure captures relative changes in duration (Methods, section \ref{M7:durdiff}). 

Thus, each subject's spatiotemporal seizure variability was described by two matrices: a pathway dissimilarity matrix (Fig. \ref{F:measures}E), containing pairwise comparisons of seizure pathways through network space, and a duration difference matrix (Fig. \ref{F:measures}F), composed of pairwise differences of seizure durations. In our subsequent analyses, we used these two measures to explore the relationship between seizure pathways and seizure durations in each subject. As such, our analysis focused on \textit{differences} in seizure pathways and durations between pairs of seizures, rather than the pathway and duration features themselves. This seizure pair approach had two main advantages. First, unlike seizure duration, seizure pathways do not map onto a single feature that changes from seizure to seizure \citep{Schroeder2020}.
However, our pairwise measures allowed us to ask questions such as, ``Does a pair of seizures have similar pathways if and only if they have similar durations?" Second, comparing these features at the seizure pair level was a more appropriate analysis for features that vary on a spectrum. In many subjects, seizures cannot be clearly grouped based on their pathways \citep{Schroeder2020} or durations \citep{Cook2016, Karoly2018b} because these features vary continuously, producing a spectrum of seizure dynamics. Thus, our pairwise approach allowed us to precisely compare seizure pathways and durations in all subjects.
\FloatBarrier
\subsection{The strength of the relationship between seizure pathways and seizure durations varies across subjects}

\begin{figure}[hbp]
	\centering
		\includegraphics{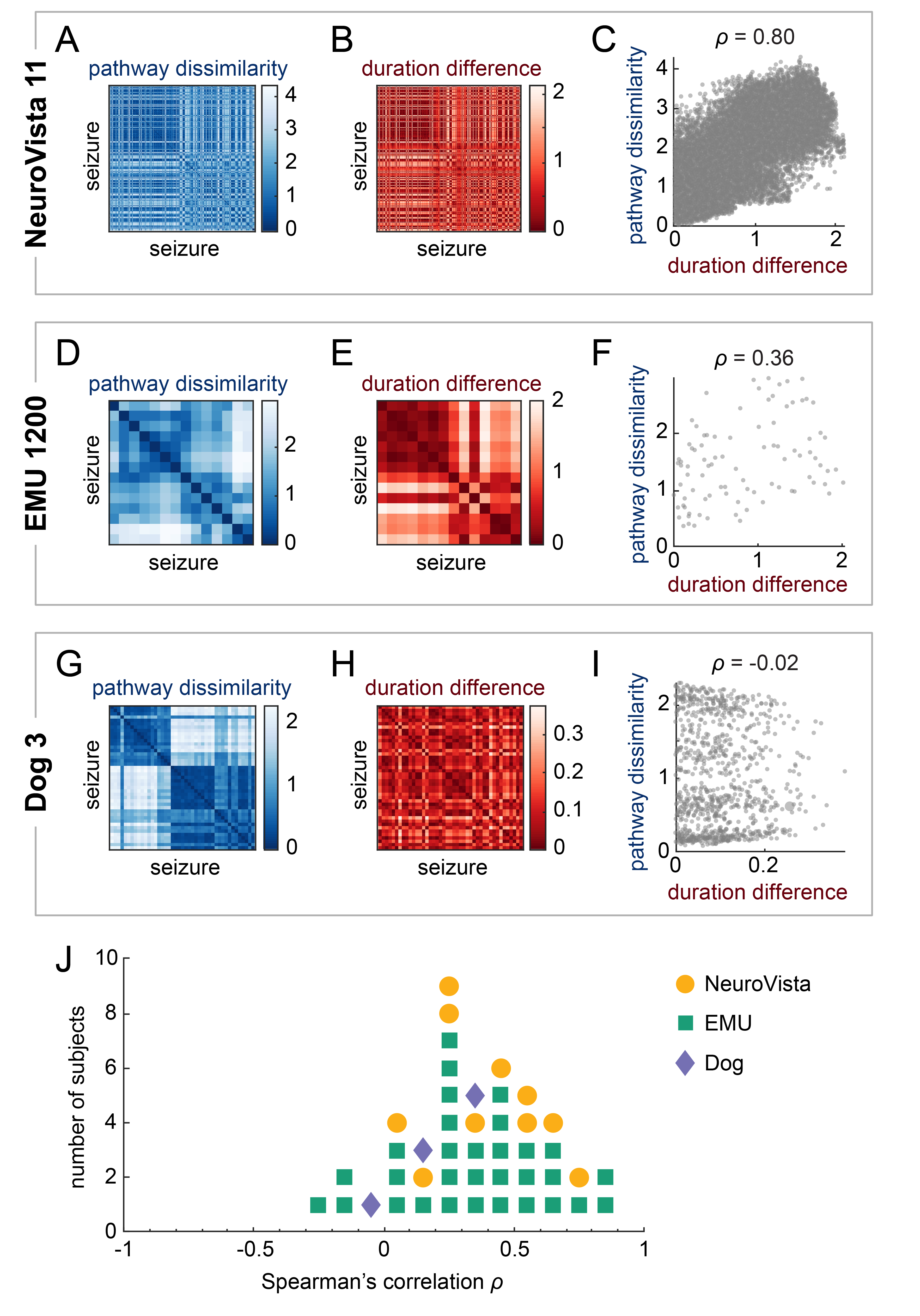}
		\caption[Comparison of pathway dissimilarities and duration differences]{\textbf{Comparison of pathway dissimilarities and duration differences.} A-I) Comparison of pathway dissimilarities and duration differences in three example subjects: NeuroVista 11 (351 seizures), EMU 1200 (14 seizures), and Dog 3 (43 seizures). A,D,G) Pathway dissimilarity matrices of the example subjects. Each matrix quantifies the pairwise dissimilarities of the subject's seizure pathways. B,E,H) Duration difference matrices in the same subjects. Each matrix quantifies the pairwise differences in the subject's seizure durations on a natural logarithm scale. C,F,I) Scatter plots and Spearman's correlations of each subject's pathway dissimilarities vs. duration differences. Each point corresponds to a seizure pair. J) Dot plot of the Spearman's correlations between pathway dissimilarities and duration differences of all subjects. Each marker corresponds to a subject, with the colour and shape indicating the subject's cohort.}
		\label{F:rho}
\end{figure}

We first compared each subject's pathway dissimilarity matrix to their duration difference matrix. Fig.~\ref{F:rho} shows the matrices of three example subjects, one from each cohort. Visually and quantitatively comparing the matrices within each subject revealed that their concordance varied across subjects. NeuroVista 11's pathway dissimilarity (Fig. \ref{F:rho}A) and duration difference (Fig. \ref{F:rho}B) matrices had very similar structures, indicating that seizures with similar (dissimilar) pathways also had similar (different) durations. The correlation between these two matrices was, as expected, very high (Fig.~\ref{F:rho}C, $\rho=0.80$).  On the other hand, Dog 3's matrices (Fig. \ref{F:rho}G,H,I, $\rho=-0.02$) had  different structures, suggesting little or no relationship between pathway dissimilarity and duration differences. EMU 1200's matrices (Fig.~\ref{F:rho}D,E,F, $\rho=0.36$) were between these two extremes: while there were some similarities across the two matrices, each matrix also had distinct patterns. These examples demonstrate that the relationship between seizure pathways and seizure durations differed across subjects.

In most subjects, pathway dissimilarities and duration differences were weakly to moderately correlated (Fig. \ref{F:rho}J, median correlation: 0.322, first quartile: 0.191, third quartile:  0.537). \ref{S6:rho_sig} provides additional information on the statistical significance of these associations, and  \ref{S6:rho_vs_max} shows that the association strength is also not determined by the range in either feature. The weak to moderate correlations revealed that changes in seizure durations were not fully explained by changes in seizure pathways and \textit{vice versa}. Therefore, seizure pathways and durations contained complementary information about the dynamics of a given seizure.

\FloatBarrier
\subsection{The relationship between pathways and durations is strengthened by pairs of seizures with both similar, or both dissimilar, pathways and durations}

We next examined how pairwise relationships between seizures could strengthen or weaken the association between seizure pathways and seizure duration within each subject. A pair of seizures could fall into one of four possible categories:

\begin{enumerate}
	\item The seizure pair had similar pathways and similar durations.
	\item The seizure pair had different pathways and different durations.
	\item The seizure pair had similar pathways, but different durations.
	\item The seizure pair had different pathways, but similar durations.
\end{enumerate}

\begin{figure}[hbp]
	\centering
		\includegraphics[width=0.9\textwidth]{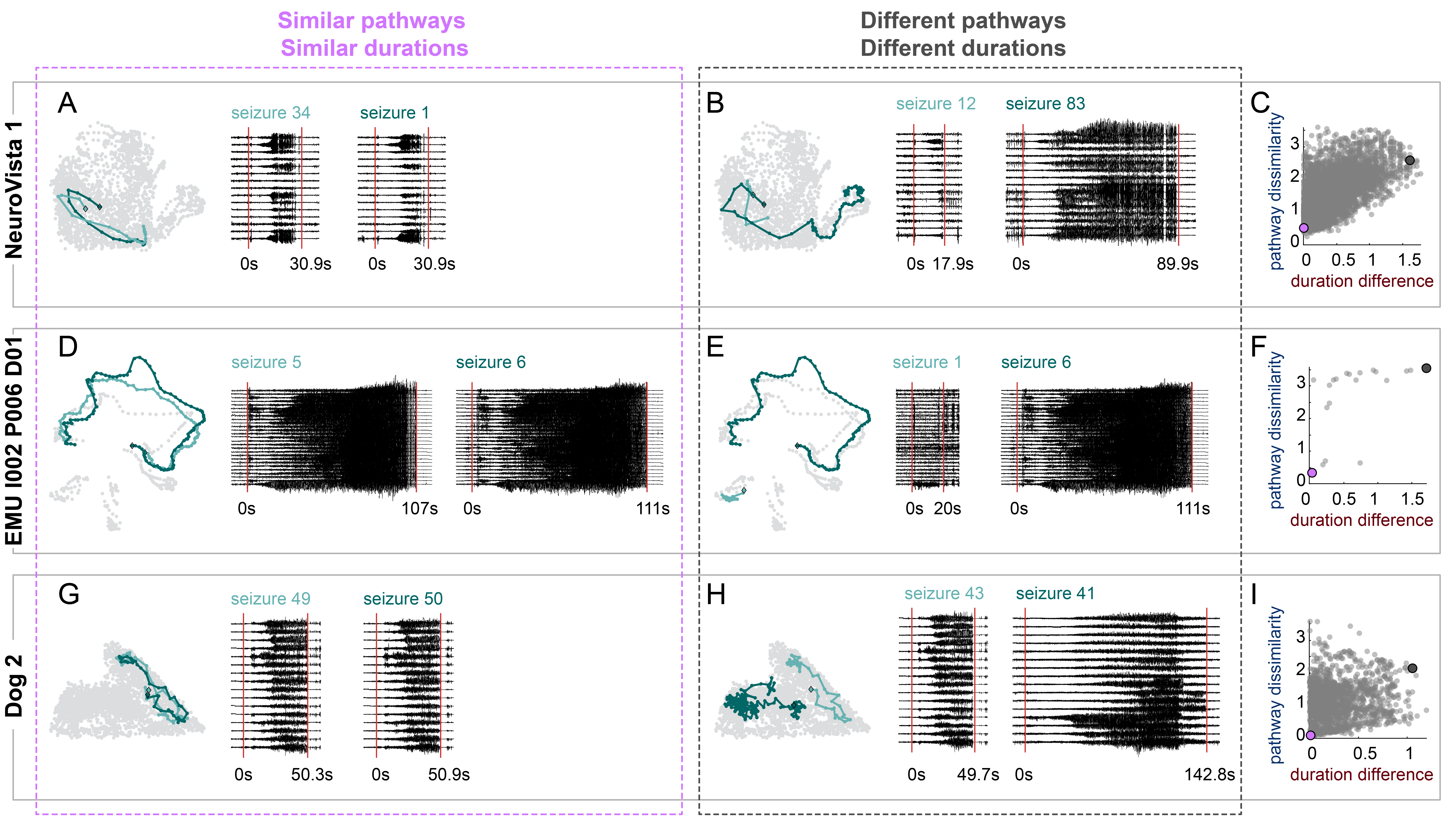}
		\caption[Example seizure pairs that strengthen the relationship between seizure pathways and seizure durations]{\textbf{Example seizure pairs that strengthen the relationship between seizure pathways and seizure durations.} A,D,G) Example pairs of seizures with similar pathways and similar durations. Left: the seizures' pathways (light teal and dark teal), embedded in network space as in Fig. \ref{F:measures}C. The time points in the subjects other seizure pathways are shown for reference (light grey points). Right: The \ac{iEEG} traces and durations of each pair of seizures, with 10s of preictal and postictal data also shown. Red lines mark seizure onset and termination. For EMU I002 P006 D01, a representative subset of channels is shown. B,E,H) For the same subjects as in A,D, and G, example pairs of seizures with different pathways and different durations. Visualisation formats are the same as in A,D, and G. For A,B,D,E,G, and H, the time and voltage scales of the \ac{iEEG} traces are consistent for each subject, but not across subjects. C,F,I) Scatter plots of pathway dissimilarities vs. duration differences of the three example subjects, with the example seizure pairs highlighted in purple (similar pathways, similar durations) and dark grey (different pathways, different durations).}
		\label{F:strengthen_pairs}
\end{figure}

We initially evaluated cases in which the seizure pair's pathway and duration agreement was concordant (i.e., both features similar or both features different, cases 1 and 2). Fig. \ref{F:strengthen_pairs} shows example pairs of seizures that had similar pathways and similar durations (case 1, Fig. \ref{F:strengthen_pairs}A,D,G) or different pathways and different durations (case 2, Fig. \ref{F:strengthen_pairs}B,E,H). In the latter case, pathways could either partially overlap in network space (Fig. \ref{F:strengthen_pairs}B) or occupy distinct regions (Fig. \ref{F:strengthen_pairs}E,H). Therefore, these disparate pathways could either share network features or have completely unrelated evolutions.

Fig. \ref{F:strengthen_pairs}C,F,I, visualises how these pairs of seizures impact the relationship between pathway and duration variability in each of the example subjects. When pathways and durations were similar, the seizure pair had a low pathway dissimilarity and a low duration difference (purple points). In contrast, pairs of seizures with different pathways and durations had high pathway dissimilarities and high duration differences (dark grey points). The combination of such seizure pairs within the same subject contributed to the positive correlations between pathway dissimilarities and duration differences. Coinciding intermediate levels of changes in both seizure pathways and seizure durations also strengthened this relationship. In other words, pathway and duration variability were related when changes in pathways produced proportional changes in durations and \textit{vice versa}.
\FloatBarrier
\subsection{The relationship between pathways and durations is weakened by elastic pathways and duplicate durations}

 \begin{figure}[hbp]
	\centering
		\includegraphics[width = 0.6\textwidth]{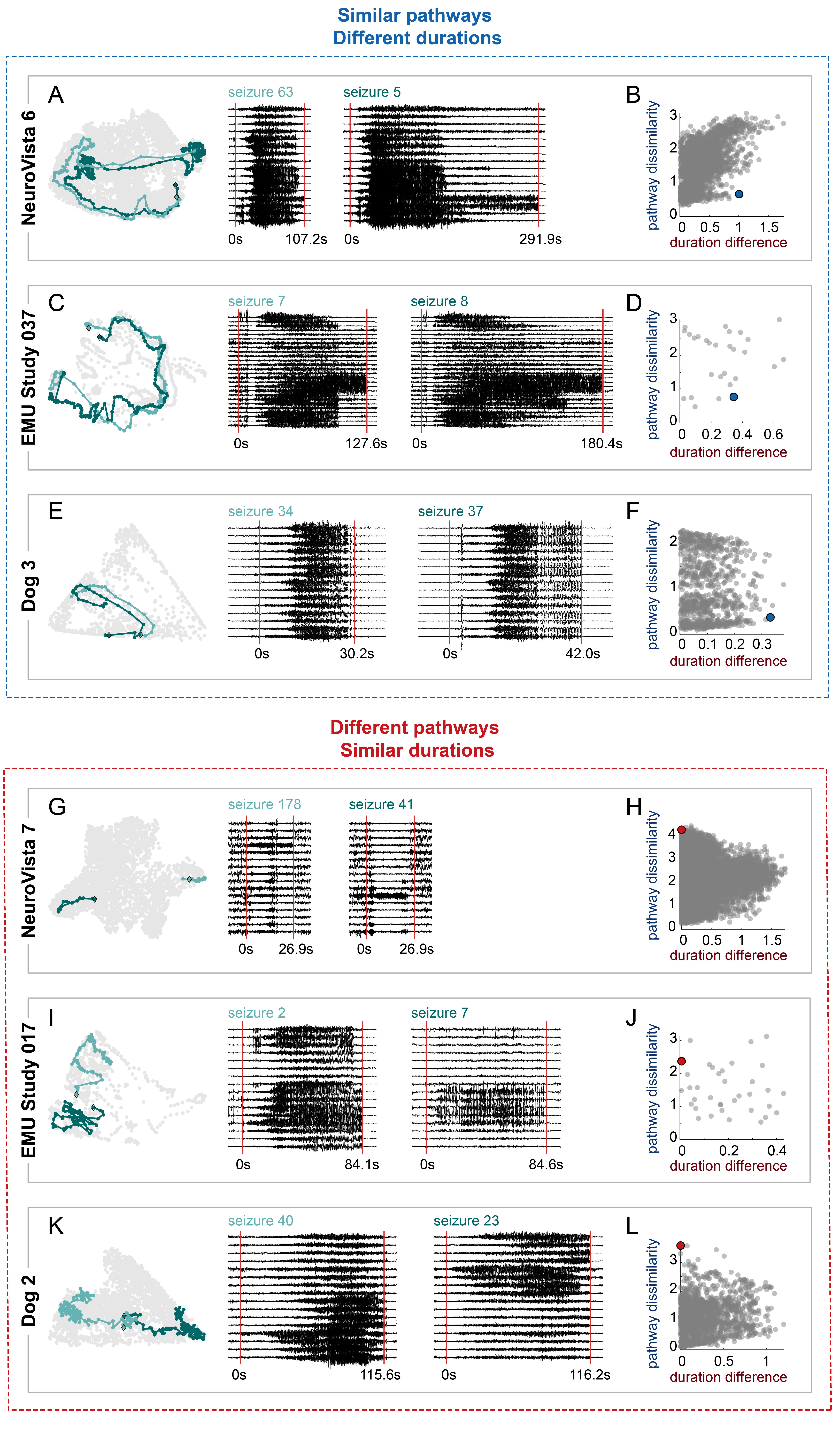}
		\caption[Example seizure pairs that weaken the relationship between seizure pathways and seizure durations]{\textbf{Example seizure pairs that weaken the relationship between seizure pathways and seizure durations.} A-F) Examples of seizure pairs with similar pathways and different durations (``elastic pathways"). A,C,E) Visualisation of the seizure pathways, durations and \ac{iEEG} is the same as in Fig. \ref{F:strengthen_pairs}. For EMU Study 037, a representative subset of channels is shown. B,D,F) Scatter plots of pathway dissimilarities vs. duration differences for each subject, with the example seizure pairs highlighted in blue. G-L) Examples of seizure pairs with different pathways and similar durations (``duplicate durations"). G,I,K) Visualisation of the seizure pathways, durations and \ac{iEEG}, as in Fig. \ref{F:strengthen_pairs}. H,J,L) Scatter plots of pathway dissimilarities vs. duration differences for each subject, with the example seizure pairs highlighted in red.}
		\label{F:weaken_pairs}
\end{figure}

We next examined how pairs of seizures could weaken the relationship between seizure pathways and durations. First, a pair of seizures could have similar pathways, but different durations (case 3). Fig \ref{F:weaken_pairs}A-F provides three examples of this scenario. Although the seizures in each pair followed similar routes through network space, they took different amounts of time to do so, revealing ``elasticity" in each example seizure pathway \citep{Wenzel2017}. Interestingly, the pathways were not uniformly elastic; instead, there appeared to be pathway-specific locations where a pathway dwelled for different amounts of time. For example, in NeuroVista 6, seizure 5 spent relatively more time in the middle and end of the pathway (Fig.~\ref{F:weaken_pairs}A).
Due to their shared pathways and different durations, such pairs weakened (Fig.~\ref{F:weaken_pairs}B,D,F blue points) the relationship between pathways and durations. These results revealed that a seizure's duration is not rigidly constrained by its pathway.

The final scenario was that two seizures had different pathways, but the same duration (case 4). Fig \ref{F:weaken_pairs}G-L illustrates this case, which we termed ``duplicate durations''. Thus, the duration of a seizure does not necessarily provide information about a seizure's pathway; in each example, the seizures had near-identical durations, but different pathways. These pairs of seizures all had low duration differences and high pathway dissimilarities (Fig. \ref{F:weaken_pairs}H,J,L), again weakening the relationship between pathway and duration variability in each of these subjects.

To determine the prevalence of elastic pathways and duplicate durations, we set thresholds for whether two seizures had similar pathways and/or similar durations (Methods, section \ref{M7:elas}). 
Almost all subjects had elastic pathways (30/31 \ac{EMU}, 10/10 NeuroVista, and 3/3 dogs) and duplicate durations (27/31 \ac{EMU}, 10/10 NeuroVista, and 3/3 dogs) (\ref{S6:elastic_duplications}). Therefore, these mechanisms for independent variability in pathways and durations were widespread in our cohorts. 
\FloatBarrier
\subsection{Populations of short and long seizures do not reliably correspond to different seizure pathways} \label{R7:durpop}

In the previous sections, we analysed the relationship between seizure pathways and seizure durations in all subjects, regardless of the nature of their seizure dynamics. It is possible that pathways and durations are more closely related in subjects whose seizures can be grouped into distinct duration populations of short and long seizures. In particular, previous studies have hypothesised that duration populations correspond to different seizure pathways \citep{Cook2016,Karoly2018b}. 

\begin{figure}
	\centering
		\includegraphics[width=\textwidth]{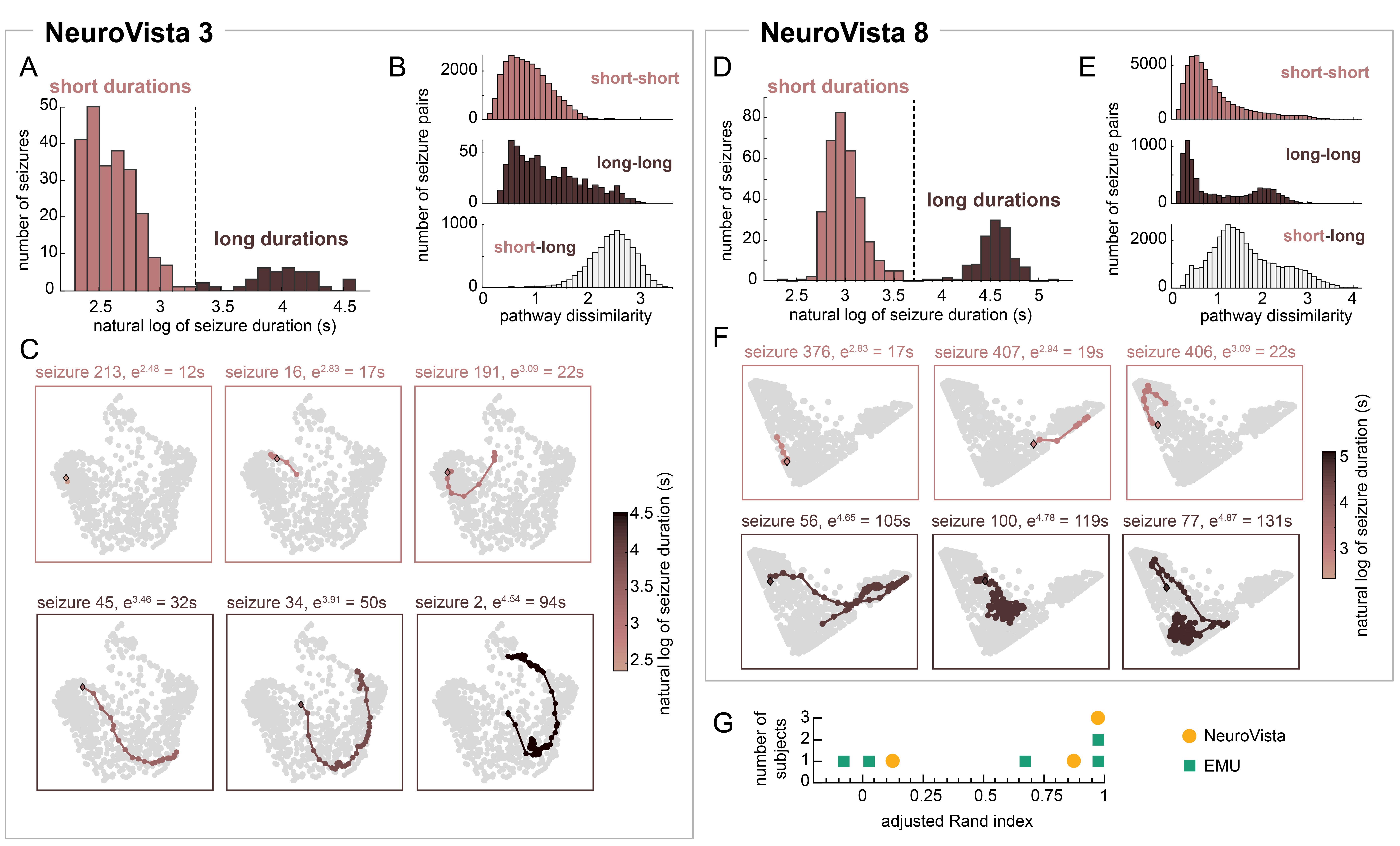}
		\caption[Short and long seizures do not necessarily correspond to different seizure pathways]{\textbf{Short and long seizures do not necessarily correspond to different seizure pathways.} Analysis of seizure pathways within subjects with multiple duration populations. A,D) Distribution of seizure durations in example subjects NeuroVista 3 (A) and NeuroVista 8 (D). Each bimodal distribution can be divided into two duration populations: one with short seizures, and one with long seizures. B,E) The distributions of pathway dissimilarities between short-short (top), long-long (middle), and short-long (bottom) pairs of seizures in each example subject. C,F) Example seizure pathways of short and long seizures in each subject. G) The adjusted Rand index between seizure duration populations and and seizure pathway clusters in all subjects with duration populations.}
		\label{F:dur_populations}
\end{figure}

As in previous work \citep{Cook2016,Karoly2018b}, we clustered seizure durations in each subject and found those subjects with multiple groups, or populations, of seizures based on their seizure durations. While most subjects did not have multiple duration populations, a total of eight subjects (5/31 \ac{EMU} patients, 3/10 Neurovista patients, and 0/3 dogs) had two duration populations. Fig. \ref{F:dur_populations}A-F explores the relationship between these duration populations and the corresponding seizure pathways in two example subjects, NeuroVista 3 and NeuroVista 8. In NeuroVista 3, pairs of seizures tended to have similar pathways if and only if they belonged to the same duration population (i.e., if they were both short or both long) (Fig. \ref{F:dur_populations}B,C). Although there was still some pathway variability within each duration population, especially among the long seizures, overall the different duration populations corresponded to different seizure pathways. In contrast, in NeuroVista 8, pairs of seizures with different durations often had more similar pathways than pairs of seizures with similar durations (Fig. \ref{F:dur_populations}E). 
Seizures with similar durations could occupy different parts of network space, while seizures with different durations (for example, short seizure 407 and long seizure 56) could partially overlap in network space (Fig. \ref{F:dur_populations}F). As a result, seizure duration populations did not distinguish different seizure pathways in NeuroVista 8.

To quantify the agreement between seizure pathways and durations populations, we also clustered seizures into two groups based on seizure dissimilarities. We compared these pathway group assignments to duration populations using the \ac{ARI}  (Fig. \ref{F:dur_populations}G). An \ac{ARI} of one indicated perfect agreement between the two partitions, while an \ac{ARI} close to zero corresponded to only chance levels of agreement. NeuroVista 3 was one of three subjects with an \ac{ARI} of one, indicating that short and long seizures perfectly corresponded to the division of seizure pathways. Meanwhile, NeuroVista 8's \ac{ARI} was only 0.14; in this subject and two others, short and long seizures were not proxies for different seizure pathways. The remaining two subjects had intermediate levels of agreement between pathway groups and duration populations. 
\ref{S6:rand_sig} contains additional clustering analyses. These results revealed a complex, subject-specific relationship between seizure durations and pathways in subjects with multiple duration populations. In some subjects, duration populations indeed corresponded to different seizure pathways, although there was additional pathway variability within each duration population. In others, duration populations were not associated with different groups of seizure pathways. 

\subsection{Relationship to other clinical variables}
Finally, we investigated if seizure pathway variability, seizure duration variability, or their relationship was associated with clinical variables, such as disease duration or lobe of epilepsy, in \ref{M7:clin}. We could confirm some previously reported relationships with seizure duration, but found no other strong relationships. Specifically, the subject-specific relationship between seizure pathways and durations was not explained by our clinical variables.

\section{Discussion}

We quantitatively compared two features of seizure dynamics: seizure durations and seizure pathways. We found that these features often varied independently within individual subjects. Seizures with the same pathway could have different durations due to temporally elastic progressions, and seizures with the same duration could have different pathways. Additionally, we found that the level of association between pathways and durations was subject-specific: in some subjects, seizures with similar pathways usually had similar durations, while in others, there was little to no association between pathway and duration similarity. Even in subjects with distinct populations of short and long seizures, seizures with different durations did not necessarily correspond to different seizure pathways. Thus, seizure pathways and durations carry complementary information about seizure dynamics, and these features can perhaps be modulated independently within a given subject. Additionally, the highly subject-specific relationship between seizure pathways and durations highlights the need for statistical and dynamical models that can be tailored to individual data. 

The fact that each individual seizure pathway did not have a rigidly predetermined duration indicates that a seizure's evolution itself does not fully dictate its rate of progression. 
More broadly, a seizure is not a pre-programmed sequence of pathological electrographic patterns with set timings. Rather, both the seizure's pattern of activity and the timings of those different patterns can change from one seizure to the next. These observations imply that there are factors that modulate the timings within seizure pathways.  Our work here has highlighted that these modulators can impact seizure pathways independently of the seizure duration, suggesting that there are multiple biological mechanisms that influence seizure features. Identifying these mechanisms could offer therapeutic targets for controlling seizures.

One potential therapeutic approach would be to reduce the duration of a given seizure pathway. To achieve this goal, further work is needed to determine the mechanisms that produce changes in seizure duration among seizures with similar pathways. In a rodent model, \cite{Wenzel2017} observed seizures with similar propagation patterns, but different durations, which they described as ``elasticity" of seizure propagation. To our knowledge, seizure elasticity has not been quantitatively described in humans before, although seizures with consistent neuronal spiking patterns, but different durations, have been observed \citep{Truccolo2011}. Our work reveals that temporal elasticity is also a common feature of human seizures. Interestingly, it appeared that such elasticity did not necessarily affect the entire seizure pathway; instead, a seizure could selectively dwell in certain parts of a given pathway. Further research is needed to understand what parts of seizure pathways are most prone to variable rates of progression as well as the underlying molecular mechanisms, such as local \citep{Wenzel2017} or feedforward \citep{Trevelyan2007} inhibition, that determine these temporal features.
Uncovering these mechanisms could provide possible clinical strategies for controlling seizure progression and duration, thus reducing seizure severity. 

It is also likely that some seizure modulators affect both seizure pathways and durations, creating seizures that have both different pathways and different durations. One possibility is that these seizures could arise from truncating seizure pathways. 
In support of this hypothesis, we observed that in the same patient, seizures with short and long durations can have overlapping pathways (e.g., Fig. \ref{F:strengthen_pairs}B). An earlier modelling study found that variability in a patient's seizure duration was associated with seizure terminations, but not onsets \citep{Karoly2018b}. While this work used a different computational approach to characterise seizure pathways, these results suggest that seizures with different durations can share the same initial evolution. Additionally, microelectrode recordings have revealed that some patients have shorter seizures that terminate earlier along the patient's characteristic seizure evolution, again suggesting that shorter seizures can arise by truncating seizure pathways \citep{Wagner2015}. This truncation mechanism could potentially be exploited clinically to induce early termination of seizure pathways and thereby reduce seizure duration and severity. 

While clinical factors such as seizure localisation are known to impact seizure durations across patients \citep{Afra2008,Kim2011dur}, the factors that could independently modulate seizure durations within the same patient are unknown. We previously hypothesised that preictal variability in brain dynamics \citep{Schroeder2020} or continuous fluctuations in interictal brain dynamics \citep{Panagiotopoulou2021} could produce changes in seizure pathways. Likewise, fluctuations in interictal dynamics such as levels of cortical excitability and inhibition \citep{Meisel2015,Badawy2012,Enatsu2012} or functional networks \citep{Mitsis2020} could potentially affect seizure duration. As noted earlier, the observed independent variability in seizure pathways and seizure durations suggests that these features can be modulated separately in some patients. Therefore, some factors may solely shape seizure spread, patterns, and connectivity, while others could determine the rate of seizure evolution. While factors such as sleep state \citep{Bazil1997} and temporal seizure clusters \citep{Kim2011dur} are known to impact seizure duration, it is unclear whether these duration changes occur as a consequence of coinciding changes in seizure pathways. Indeed, in a rodent model, seizure durations, severity, and spread all change over the course of a seizure cluster \citep{Kudlacek2021}, suggesting that some mechanisms influence both seizure pathways and durations. Disentangling the factors that shape seizure pathways and durations would require accounting for variability in one feature when analysing the other aspect of seizure dynamics. 

One promising avenue to understanding seizure pathways and their durations is developing computational models that capture not only specific stages of a seizure, but also neural dynamics throughout the seizure's evolution. Many studies have focused on computationally analysing \citep{Salami2020,Salami2021,Alarcon1995,Afra2015,Jimenez-Jimenez2015,Lagarde2019} or modelling \citep{Jirsa2014,Saggio2020,Wang2017,Proix2018} variability in seizure onset and termination dynamics. For example, \cite{Jirsa2014} characterised seizures by the types of dynamic transitions that occurred at seizure start and end. Using this approach, \cite{Saggio2020} classified seizures into ``dynamotypes" and uncovered within-patient variability in these classes. While their classification characterises seizure onset and termination, \cite{Saggio2020} also developed a model that explains relationships between dynamotypes as well as more complex seizure dynamics such as status epilepticus. Such models can therefore also capture the seizure's full evolution, or dynamical pathway  \citep{Karoly2018b,Wendling2002}. Thus, computational models of seizure evolutions could be extended to explore the dynamical mechanisms underlying other types of seizure variability beyond seizure transitions, such as the variability in seizure pathways, dwell sites, and overall duration that we observed in this work.

Our study was limited to human patients with drug-resistant focal epilepsies and dogs with focal-onset seizures, and it is unclear whether similar relationships between seizure pathways and durations exist in other types of epilepsies. Our concordant findings in dogs indicate that our results generalise beyond human patients. It is also likely that we did not observe all types and combinations of pathway and duration variability in our subjects, especially in \ac{EMU} patients with shorter recordings \citep{King-Stephens2015}. Another limitation of our study is that seizure duration depends on clinically or algorithmically marked seizure onsets and terminations. Clinical markings can be subjective and vary from marker to marker, especially in some seizures with more ambiguous onsets \citep{Davis2018, Halford2015}. However, marking errors were likely small and non-systematic relative to the length of most seizures.

We have shown that seizure pathways and durations can vary independently within the same patient, increasing the possible combinations of seizure dynamics that can occur in a given patient. As such, both pathway and duration information is needed to fully characterise a seizure's dynamics. Determining the mechanisms by which each feature independently and co-varies could potentially identify strategies for reducing seizure duration and severity in therapeutic interventions.

\section{Methods}

\subsection{Subjects and seizure data} This work analysed seizures from three cohorts of subjects:
\begin{enumerate}
    \item \textbf{\Ac{EMU} patients:} 31 patients with refractory focal epilepsy whose continuous \ac{iEEG} recordings were acquired during presurgical evaluation at the Mayo Clinic (MC) (12 patients), the Hospital of the University of Pennsylvania (HUP) (1 patient), or the University College London Hospital (UCLH) (18 patients). MC and HUP patient data is available on the IEEG Portal, www.ieeg.org \citep{Wagenaar2013, Kini2016}, and all IEEG Portal patients gave consent for their anonymised \ac{iEEG} to be available through this database. The \ac{iEEG} recordings of the UCLH patients were anonymised, exported, and analysed under the approval of the Newcastle University Ethics Committee (reference number 6887/2018). The same \ac{EMU} cohort and seizure data was used in our previous analysis of seizure variability \citep{Schroeder2020}. Additional information about each patient and the analysed seizures is shown in Supplementary Table \ref{emu_metadata}.
    \item \textbf{NeuroVista patients:} Seizures from 10 NeuroVista patients were included to analyse seizure variability over longer timescales in patients with focal epilepsy. The NeuroVista seizure \ac{iEEG} data from \cite{Karoly2018b}, which includes 12 patients, was used for this analysis. Patients NeuroVista 2 and NeuroVista 4 were removed from our analysis due to low numbers of analysable seizures. The patients and collection of their chronic \ac{iEEG} data is described in detail in \cite{Cook2013}. Briefly, all patients had refractory focal epilepsy and experienced 2-12 seizures per month. For the NeuroVista seizure prediction study, each patient was implanted with 16 surface \ac{iEEG} electrodes over the brain quadrant thought to contain the epileptogenic zone. Additional patient details are provided in Supplementary Table \ref{nv_metadata}.
    \item \textbf{Dogs:} To explore seizure variability over longer timescales and in non-human subjects, \ac{iEEG} was also analysed from three dogs with focal-onset seizures due to naturally occurring canine epilepsy. The dogs underwent prolonged recordings to test a novel implantable electrographic recording device \citep{Howbert2014,Davis2011} (recording data available on the IEEG Portal, www.ieeg.org \citep{Wagenaar2013, Kini2016}). 
    Metadata for the dogs is provided in Supplementary Table \ref{canine_metadata}.
\end{enumerate}

For all cohorts, seizures were required to have clear electrographic correlates and durations of at least 10s. We excluded seizures with noisy or large missing segments. Subclinical seizures were included in the analysis. For NeuroVista patients, seizures with clinical manifestations and corresponding \ac{iEEG} changes (referred to as ``type 1" seizures in previous literature \citep{Cook2016,Karoly2018b}) and seizures with \ac{iEEG} changes comparable to type 1 seizures, but without confirmed clinical manifestations (previously referred to as ``type 2" seizures) were included in the analysis. 

For \ac{EMU} patients, seizure onsets and terminations were marked by the corresponding clinical teams. For NeuroVista patients, seizure onsets and terminations were marked by clinical staff, with seizure detection and classification aided by using patient diaries, audio recordings, and a seizure detection algorithm \citep{Cook2016}. For dogs, seizure onsets were marked by expert readers and an algorithm was used to detect seizure termination (see \ref{SS:canine_termination}). 

\subsection{Intracranial EEG preprocessing for epilepsy monitoring unit patients and dogs} \label{SS:preprocessing}
For each subject, if different seizures were recorded at multiple sampling frequencies, all of the recordings were first downsampled to the lowest sampling frequency. Noisy channels were removed based on visual inspection and short missing segments ($<$0.05s, with the exception of one 0.514s segment in patient ``Study 020”) were linearly interpolated. All channels were re-referenced to a common average reference. Each channel’s time series was then bandpass filtered from 1-150 Hz (4th order, zero-phase Butterworth filter) and notch filtered (4th order, 2 Hz width, zero-phase Butterworth filter) at 60 and 120 Hz (IEEG Portal patients and dogs) or 50, 100, and 150 Hz (UCLH patients).

\subsection{Seizure \ac{iEEG} preprocessing for NeuroVista patients} \label{SS:NV_preprocessing}

NeuroVista seizure data was previously notch filtered at 50 Hz during the \ac{iEEG} acquisition and bandpass filtered (2nd order, zero-phase Butterworth filter from 1-180 Hz) by \cite{Karoly2018b}. We then removed any electrodes with noisy or intermittent signal from the seizure analysis and re-referenced all \ac{iEEG} to a common average reference.

The NeuroVista data contains time periods of signal dropouts when the \ac{iEEG} signal was not recorded. We detected and removed periods of signal dropout by using line length to identify \ac{iEEG} segments with no signal (i.e., a flat time series with no voltage changes). We defined the line length $L$ of a time series as
\begin{center}
$L =  \dfrac{1}{T-1} \sum_{i=1}^{T - 1} | x_{i+1} - x_i | $
\end{center}
where $x_i$ is the $i$th time point in a time series with $T$ time points. 

For each seizure, time-varying line length was computed for each \ac{iEEG} channel in sliding windows (1/10s window, 1/20s overlap). If a time window had at least 8 out of 16 channels with line length $\leq$ 0.5, that time window along with the preceding and following time windows were considered missing data. Section \ref{SS:fc} describes how this missing data was handled during the computation of seizure time-varying functional connectivity.

\subsection{Computing seizure time-varying functional connectivity} \label{SS:fc}

The time-varying functional connectivity, defined as coherence in six frequency bands (delta 1-4 Hz, theta 4-8 Hz, alpha 8-13 Hz, beta 13-30 Hz, gamma 30-80 Hz, high gamma 80-150 Hz), was computed for each seizure, from the marked seizure onset to termination, using a 10s sliding window with 9s overlap between consecutive windows, as in previous work \citep{Schroeder2020}. Coherence in each frequency band was computed using band-averaged coherence, defined as
\begin{center}
	$C_{i,j} = \dfrac{|\sum_{f=f_1}^{f_2} P_{i,j}(f)|^2}{\sum_{f=f_1}^{f_2} P_{i,i}(f) \sum_{f=f_1}^{f_2} P_{j,j}(f)}$
\end{center}

where $f_1$ and $f_2$ are the lower and upper bounds of the frequency band, $P_{i,j}(f)$ is the cross-spectrum density of channels $i$ and $j$, and $P_{i,i}(f)$ and $P_{j,j}(f)$ are the autospectrum densities of channels $i$ and $j$, respectively. For each 10s window, auto-spectra and cross-spectra were calculated using Welch’s method (2s sliding window with 1s overlap). As noted in section \ref{SS:NV_preprocessing}, many seizures in NeuroVista patients contained missing data due to signal dropouts. We tolerated some missing data in this cohort by allowing functional connectivity in each 10s window to be estimated using a subset of the 2s Welch subwindows. Specifically, for each functional connectivity time window, we only treated the 10s functional connectivity time window as missing data if five or more of the 2s subwindows contained missing data. Any NeuroVista seizures with missing functional connectivity time windows were excluded from the remainder of the analysis. 

Each resulting coherence matrix was re-expressed in vector form by re-arranging the upper-triangular, off-diagonal elements into a vector of length $(n^2 - n)/2$, where $n$ is the number of \ac{iEEG} channels, and the vector was normalised to have an $L1$ norm of 1. Each seizure time window was therefore represented by a total of $6 \times (n^2 - n)/2$ features that captured the pairwise channel interactions in the six different frequency bands. In a seizure with $m$ time windows, the seizure's time-varying functional connectivity was described by a multivariate time series with $6 \times (n^2 - n)/2$ features and $m$ time points.

To reduce noise in the connectivity matrices, patterns of recurring functional connectivity were extracted in each subject using stability \ac{NMF} \citep{Lee1999,Wu2016} using the same pipeline as in our previous work \citep{Schroeder2020}. The \ac{NMF} decomposition was used to reconstruct a low-rank approximation of the seizure functional connectivity time series that was used for all downstream analysis. The time-varying functional connectivity of each seizure was also referred to as ``seizure network evolutions" and ``seizure pathways" throughout the text.

\subsection{Visualising seizure pathways} 

To visualise changes in seizure networks within and between seizures, each subject's seizure time-varying functional connectivity (``seizure pathway") was projected into two-dimensional space using Sammon mapping, a variation of \ac{MDS} \citep{Sammon1969}. The mapping approximated the $L1$ distances between the functional connectivity patterns of each pair of seizure time windows such that seizure time windows with more similar functional connectivity were placed closer together in the projection.

\subsection{Comparing seizure pathways using pathway dissimilarities}

We used the approach of \cite{Schroeder2020} to compare pairs of seizure pathways, which were  described by the seizure time-varying functional connectivity, within each patient. Briefly, for each pair of seizures, we used \ac{DTW} \citep{Sakoe1978} (MATLAB function \textit{dtw}) to align similar time points in their functional connectivity time series and minimise the overall $L1$ distance between the pair of time series. The seizure pair's ``pathway dissimilarity" (previously called ``seizure dissimilarity" in \cite{Schroeder2020}) was then defined as the average $L1$ distance between their warped time series. Repeating this process for each pair of a subject's $s$ seizures yielded the subject's pathway dissimilarity matrix, a symmetric $s \times s$ matrix containing all of the pairwise pathway dissimilarities.  

\subsection{Comparing seizure durations using duration differences} \label{M7:durdiff}

To compare seizure durations, we computed a pairwise ``duration difference" measure for each pair of a subject's seizures. First, as in previous work \citep{Cook2016}, we transformed each subject's seizure durations by computing their natural logarithm, which made each subject's distribution of seizure durations closer to a normal distribution. We then defined the duration difference between a pair of seizures as the absolute difference between their transformed durations, 
\begin{center}
	$|ln(l_i) - ln(l_j)|$
\end{center}
where $l_i$ and $l_j$ are the durations, in seconds, of seizures $i$ and $j$, respectively. Due to the properties of logarithms, this measure is equal to 
\begin{center}
	$|ln(l_i/l_j)|$
\end{center}
and therefore depends on the ratio between the durations of seizures $i$ and $j$. As such, duration differences capture the proportional differences between seizure durations.
For example, the duration difference between a 20s seizure and a 40s seizure will be the same as the duration difference between a 60s seizure and a 120s because in both cases, the longer seizure is twice the duration of the shorter seizure. Likewise, a certain absolute change in duration, such as 10s, results in a larger duration difference when the original seizure is shorter. As for the pathway dissimilarity measure, duration differences were computed for each pair of a subject's $s$ seizures to create the subject's symmetric $s \times s$ duration difference matrix.

\subsection{Comparing pathway dissimilarities and duration differences}

To compare pathway dissimilarities and duration differences, we used the same approach as \cite{Schroeder2020} for comparing dissimilarity matrices. For each subject, we computed Spearman's correlation between the upper triangular elements of their pathway dissimilarity and duration difference matrices. We then used the Mantel test \citep{Mantel1967} (10,000 permutations, one-sided significance test) to determine the probability of obtaining a correlation greater than or equal to the observed correlation by chance. 

\subsection{Defining elastic pathways and duplicate durations} \label{M7:elas}

To determine the prevalence of elastic pathways and duplicate durations, we set first thresholds for defining whether a seizure pair had similar pathways (pathway dissimilarity $\leq 1$) and similar durations (duration difference $\leq 0.2$). The pathway threshold was chosen because seizures with pathway dissimilarities below the threshold tend to have visually similar pathways and electrographic patterns. The duration difference threshold allows a  $e^{0.2} ~= 1.22$ fold increase in duration relative to the shorter seizure before the durations are considered different. These thresholds were set so that the overall proportion, across all subjects, of seizure pairs with similar pathways was comparable to the proportion of seizure pairs with similar durations (32.6\% and 34.7\%, respectively). Seizure pairs with similar pathways (pathway dissimilarity  $\leq 1$) and different durations (duration difference $> 0.2$) were then defined as examples of elastic pathways. Seizure pairs with different pathways (pathway dissimilarity  $> 1$) and similar durations (duration difference $\leq 0.2$) were considered examples of duplicate durations. 

\subsection{Comparing duration populations and seizure pathways}

In each subject, we determined the number of duration populations by clustering seizure durations (after the natural logarithm transformation) using \textit{k}-means, with the number of clusters ($k$) scanned from 1 to 5. The gap statistic \citep{Tibshirani2001} (MATLAB \textit{evalclusters}, search method \textit{firstMaxSE}, with 1000 reference distributions) was used to select the optimal number of clusters, with $k = 1$ indicating an absence of multiple duration populations and $k \geq 2$ revealing multiple duration populations.

In subjects with multiple duration populations, we additionally clustered seizures into an equivalent number of groups ($k$) to compare seizure clusters based on duration with clusters based on pathways. In each subject, seizure pathways were clustered by applying \ac{UPGMA} hierarchical clustering to the pathway dissimilarity matrix, and the resulting dendrogram was then cut to produce the same number of discrete pathway clusters as duration populations. The Rand index and adjusted Rand index were then computed to compare the duration population and pathway clusters partitions. To determine the statistical significance of these measures, the cluster membership for one partition was permuted 10,000 times and the measures were recomputed for each partition to create null distributions.  

\subsection{Correction for multiple comparisons} \label{M7:fdr}

The Benjamini-Hochberg false discovery rate \ac{FDR} correction \citep{Benjamini1995}, with $\alpha = 0.05$ ($\alpha = 0.10$ to identify statistical trends),  was applied to the set of $p$-values from all statistical tests in this work and corresponding supplementary material.  
Uncorrected p-values are reported in the text, with statistical significance determined after FDR correction.

\subsection{Code and data availability}

All analyses were performed using custom code and built-in functions in MATLAB 2018b. \ac{NMF} was performed using the Nonnegative Matrix Factorization Algorithms Toolbox (https://github.com/kimjingu/nonnegfac-matlab/) \citep{Kim2014,Kim2011}. The seizure \ac{iEEG} data of the IEEG Portal \ac{EMU} patients and dogs is available at www.ieeg.org \citep{Wagenaar2013, Kini2016}. The NeuroVista seizure data is available upon request from \cite{Karoly2018b}. The decomposed seizure pathways (\ac{NMF} $W$ and $H$ matrices) and seizure durations of all subjects, along with code for the downstream analysis in the main text, will be available on Zenodo (DOI 10.5281/zenodo.5503590) upon publication. 

\section{Acknowledgements}
We thank members of the Computational Neurology, Neuroscience \& Psychiatry Lab (www.cnnp-lab.com) for discussions on the analysis and manuscript; and Catherine Scott and Roman Rodionov for helping with data organization in the UCLH dataset. B.D. receives support from the NIH National Institute of Neurological Disorders and Stroke U01-NS090407 (Center for SUDEP Research) and Epilepsy Research UK. Y.W. gratefully acknowledges funding from Wellcome Trust (208940/Z/17/Z). P.N.T. is supported by a UKRI Future Leaders Fellowship (MR/T04294X/1). 
\newpage

\section{Author contributions}
Conceptualization: GMS, YW

Formal Analysis: GMS

Resources: FAC, MJC, BD, JSD, PJK, PNT, YW

Writing and editing: GMS, FAC, MJC, BD, JSD, PJK, PNT, YW

\newpage

\bibliography{references}

\begin{thebibliography}{61}
\providecommand{\natexlab}[1]{#1}
\providecommand{\url}[1]{\texttt{#1}}
\expandafter\ifx\csname urlstyle\endcsname\relax
  \providecommand{\doi}[1]{doi: #1}\else
  \providecommand{\doi}{doi: \begingroup \urlstyle{rm}\Url}\fi

\bibitem[Afra et~al.(2008)Afra, Jouny, and Bergey]{Afra2008}
Pegah Afra, Christophe~C. Jouny, and Gregory~K. Bergey.
\newblock Duration of complex partial seizures: An intracranial eeg study.
\newblock \emph{Epilepsia}, 49\penalty0 (4):\penalty0 677--684, 2008.
\newblock \doi{https://doi.org/10.1111/j.1528-1167.2007.01420.x}.
\newblock URL
  \url{https://onlinelibrary.wiley.com/doi/abs/10.1111/j.1528-1167.2007.01420.x}.

\bibitem[Afra et~al.(2015)Afra, Jouny, and Bergey]{Afra2015}
Pegah Afra, Christopher~C. Jouny, and Gregory~K. Bergey.
\newblock {Termination patterns of complex partial seizures: An intracranial
  EEG study}.
\newblock \emph{Seizure}, 32:\penalty0 9--15, 2015.
\newblock ISSN 15322688.
\newblock \doi{10.1016/j.seizure.2015.08.004}.
\newblock URL \url{http://dx.doi.org/10.1016/j.seizure.2015.08.004}.

\bibitem[Alarcon et~al.(1995)Alarcon, Binnie, Elwes, and Polkey]{Alarcon1995}
G~Alarcon, C~D Binnie, R~D~C Elwes, and C~E Polkey.
\newblock {Power spectrum and intracranial EEG patterns at seizure onset in
  partial epilepsy}.
\newblock \emph{Electroencephalography and Clinical Neurophysiology},
  94:\penalty0 326--337, 1995.
\newblock ISSN 0013-4694.

\bibitem[Badawy et~al.(2012)Badawy, Freestone, Lai, and Cook]{Badawy2012}
R.~A.~B. Badawy, D.~R. Freestone, A.~Lai, and M.~J. Cook.
\newblock {Epilepsy: Ever-changing states of cortical excitability}.
\newblock \emph{Neuroscience}, 222:\penalty0 89--99, 2012.
\newblock ISSN 03064522.
\newblock \doi{10.1016/j.neuroscience.2012.07.015}.
\newblock URL \url{http://dx.doi.org/10.1016/j.neuroscience.2012.07.015}.

\bibitem[Barends et~al.(2020)Barends, Walstock, Botman, {De Kruif}, Claassen,
  {Van Der Wouden}, {Olde Hartman}, Dekker, and {Van Der Horst}]{Barends2020}
Hieke Barends, Ella Walstock, Femke Botman, Anja {De Kruif}, Nikki Claassen,
  Johannes~C. {Van Der Wouden}, Tim {Olde Hartman}, Joost Dekker, and Henriette
  {Van Der Horst}.
\newblock {Patients' experiences with fluctuations in persistent physical
  symptoms: A qualitative study}.
\newblock \emph{BMJ Open}, 10:\penalty0 e035833, 2020.
\newblock ISSN 20446055.
\newblock \doi{10.1136/bmjopen-2019-035833}.

\bibitem[Baud et~al.(2015)Baud, Vulliemoz, and Seeck]{Baud2015}
Maxime~O. Baud, Serge Vulliemoz, and Margitta Seeck.
\newblock {Recurrent secondary generalization in frontal lobe epilepsy:
  predictors and a potential link to surgical outcome?}
\newblock \emph{Epilepsia}, 56\penalty0 (9):\penalty0 1454--1462, 2015.
\newblock ISSN 00139580.
\newblock \doi{10.1111/epi.13086}.
\newblock URL \url{http://doi.wiley.com/10.1111/epi.13086}.

\bibitem[Bazil and Walczak(1997)]{Bazil1997}
C~W Bazil and T~S Walczak.
\newblock {Effects of sleep and sleep stage on epileptic and nonepileptic
  seizures}.
\newblock \emph{Epilepsia}, 38\penalty0 (1):\penalty0 56--62, 1997.
\newblock ISSN 0013-9580.
\newblock \doi{10.1111/j.1528-1157.1997.tb01077.x}.

\bibitem[Benjamini and Hochberg(1995)]{Benjamini1995}
Yoav Benjamini and Yosef Hochberg.
\newblock {Controlling the false discovery rate: a practical and powerful
  approach to multiple testing}.
\newblock \emph{Journal of the Royal Statistical Society: Series B
  (Methodological)}, 57\penalty0 (1):\penalty0 289--300, 1995.

\bibitem[Burns et~al.(2014)Burns, Santaniello, Yaffe, Jouny, Crone, Bergey,
  Anderson, and Sarma]{Burns2014}
Samuel~P Burns, Sabato Santaniello, Robert~B Yaffe, Christophe~C Jouny,
  Nathan~E Crone, Gregory~K Bergey, William~S Anderson, and Sridevi~V Sarma.
\newblock {Network dynamics of the brain and influence of the epileptic seizure
  onset zone}.
\newblock \emph{Proceedings of the National Academy of Sciences}, 111\penalty0
  (49):\penalty0 E5321--E5330, 2014.
\newblock \doi{10.1073/pnas.1401752111}.

\bibitem[Cook et~al.(2013)Cook, O'Brien, Berkovic, Murphy, Morokoff, Fabinyi,
  D'Souza, Yerra, Archer, Litewka, Hosking, Lightfoot, Ruedebusch, Sheffield,
  Snyder, Leyde, and Himes]{Cook2013}
Mark~J. Cook, Terence~J. O'Brien, Samuel~F. Berkovic, Michael Murphy, Andrew
  Morokoff, Gavin Fabinyi, Wendyl D'Souza, Raju Yerra, John Archer, Lucas
  Litewka, Sean Hosking, Paul Lightfoot, Vanessa Ruedebusch, W.~Douglas
  Sheffield, David Snyder, Kent Leyde, and David Himes.
\newblock {Prediction of seizure likelihood with a long-term, implanted seizure
  advisory system in patients with drug-resistant epilepsy: A first-in-man
  study}.
\newblock \emph{The Lancet Neurology}, 12:\penalty0 563--571, 2013.
\newblock ISSN 14744422.
\newblock \doi{10.1016/S1474-4422(13)70075-9}.

\bibitem[Cook et~al.(2016)Cook, Karoly, Freestone, Himes, Leyde, Berkovic,
  O'Brien, Grayden, and Boston]{Cook2016}
Mark~J. Cook, Philippa~J. Karoly, Dean~R. Freestone, David Himes, Kent Leyde,
  Samuel Berkovic, Terence O'Brien, David~B. Grayden, and Ray Boston.
\newblock {Human focal seizures are characterized by populations of fixed
  duration and interval}.
\newblock \emph{Epilepsia}, 57\penalty0 (3):\penalty0 359--368, 2016.
\newblock ISSN 15281167.
\newblock \doi{10.1111/epi.13291}.

\bibitem[Cramer and French(2001)]{Cramer2001}
Joyce~A. Cramer and Jacqueline French.
\newblock {Quantitative assessment of seizure severity for clinical trials: A
  review of approaches to seizure components}.
\newblock \emph{Epilepsia}, 42\penalty0 (1):\penalty0 119--129, 2001.
\newblock ISSN 00139580.
\newblock \doi{10.1046/j.1528-1157.2001.19400.x}.

\bibitem[Davis et~al.(2011)Davis, Sturges, Vite, Ruedebusch, Worrell, Gardner,
  Leyde, Sheffield, and Litt]{Davis2011}
Kathryn~A. Davis, Beverly~K. Sturges, Charles~H. Vite, Vanessa Ruedebusch,
  Gregory Worrell, Andrew~B. Gardner, Kent Leyde, W.~Douglas Sheffield, and
  Brian Litt.
\newblock A novel implanted device to wirelessly record and analyze continuous
  intracranial canine eeg.
\newblock \emph{Epilepsy Research}, 96\penalty0 (1):\penalty0 116--122, 2011.
\newblock ISSN 0920-1211.
\newblock \doi{https://doi.org/10.1016/j.eplepsyres.2011.05.011}.
\newblock URL
  \url{https://www.sciencedirect.com/science/article/pii/S0920121111001318}.

\bibitem[Davis et~al.(2018)Davis, Devries, Krieger, Mihaylova, Minecan, Litt,
  Wagenaar, and Stacey]{Davis2018}
Kathryn~A. Davis, Seth~P. Devries, Abba Krieger, Temenuzhka Mihaylova, Daniela
  Minecan, Brian Litt, Joost~B. Wagenaar, and William~C. Stacey.
\newblock {The effect of increased intracranial EEG sampling rates in clinical
  practice}.
\newblock \emph{Clinical Neurophysiology}, 129:\penalty0 360--367, 2018.
\newblock ISSN 18728952.
\newblock \doi{10.1016/j.clinph.2017.10.039}.

\bibitem[de~Tisi et~al.(2011)de~Tisi, Bell, Peacock, McEvoy, Harkness, Sander,
  and Duncan]{DeTisi2011}
Jane de~Tisi, Gail~S. Bell, Janet~L. Peacock, Andrew~W. McEvoy, William~Fj
  Harkness, Josemir~W. Sander, and John~S. Duncan.
\newblock {The long-term outcome of adult epilepsy surgery, patterns of seizure
  remission, and relapse: a cohort study}.
\newblock \emph{The Lancet}, 378:\penalty0 1388--1395, 2011.
\newblock ISSN 01406736.
\newblock \doi{10.1016/S0140-6736(11)60890-8}.
\newblock URL \url{http://dx.doi.org/10.1016/S0140-6736(11)60890-8}.

\bibitem[Dobesberger et~al.(2015)Dobesberger, Risti{\'{c}}, Walser,
  Kuchukhidze, Unterberger, H{\"{o}}fler, Amann, and Trinka]{Dobesberger2015}
Judith Dobesberger, Aleksandar~J. Risti{\'{c}}, Gerald Walser, Giorgi
  Kuchukhidze, Iris Unterberger, Julia H{\"{o}}fler, Edda Amann, and Eugen
  Trinka.
\newblock {Duration of focal complex, secondarily generalized tonic-clonic, and
  primarily generalized tonic-clonic seizures - A video-EEG analysis}.
\newblock \emph{Epilepsy and Behavior}, 49:\penalty0 111--117, 2015.
\newblock ISSN 15255069.
\newblock \doi{10.1016/j.yebeh.2015.03.023}.

\bibitem[Enatsu et~al.(2012)Enatsu, Jin, Elwan, Kubota, Piao, O'Connor,
  Horning, Burgess, Bingaman, and Nair]{Enatsu2012}
Rei Enatsu, Kazutaka Jin, Sherif Elwan, Yuichi Kubota, Zhe Piao, Timothy
  O'Connor, Karl Horning, Richard~C. Burgess, William Bingaman, and Dileep~R.
  Nair.
\newblock {Correlations between ictal propagation and response to electrical
  cortical stimulation: A cortico-cortical evoked potential study}.
\newblock \emph{Epilepsy Research}, 101:\penalty0 76--87, 2012.
\newblock ISSN 09201211.
\newblock \doi{10.1016/j.eplepsyres.2012.03.004}.
\newblock URL \url{http://dx.doi.org/10.1016/j.eplepsyres.2012.03.004}.

\bibitem[Farooque and Duckrow(2014)]{Farooque2014}
Pue Farooque and Robert Duckrow.
\newblock {Subclinical seizures during intracranial EEG recording: Are they
  clinically significant?}
\newblock \emph{Epilepsy Research}, 108:\penalty0 1790--1796, 2014.
\newblock ISSN 18726844.
\newblock \doi{10.1016/j.eplepsyres.2014.09.020}.

\bibitem[Fava et~al.(2003)Fava, Park, and Sonino]{Fava2003}
Giovanni~A. Fava, Seung~K. Park, and Nicoletta Sonino.
\newblock {Treatment of recurrent depression}.
\newblock \emph{Expert Review of Neurotherapeutics}, 17\penalty0 (15):\penalty0
  1109--1117, 2003.
\newblock ISSN 14737175.
\newblock \doi{10.1586/14737175.6.11.1735}.

\bibitem[Fisher et~al.(2017)Fisher, Cross, French, Higurashi, Hirsh, Jansen,
  Lagae, Mosh{\'{e}}, Peltola, Perez, Scheffer, and Zuberi]{Fisher2017}
Robert~S Fisher, J~Helen Cross, Jacqueline~A French, Norimichi Higurashi,
  Edouard Hirsh, Floor~E. Jansen, Lieven Lagae, Solomon~L. Mosh{\'{e}}, Jukka
  Peltola, Eliane~Roulet Perez, Ingrid~E Scheffer, and Sameer~M Zuberi.
\newblock {Operational classification of seizure types by the International
  League Against Epilepsy: position paper of the ILAE Commission for
  Classification and Terminology}.
\newblock \emph{Epilepsia}, 58\penalty0 (4):\penalty0 522--530, 2017.
\newblock \doi{10.1111/epi.13670}.

\bibitem[Frey et~al.(2011)Frey, Maksym, and Suki]{Frey2011}
Urs Frey, Geoffrey Maksym, and B{\'{e}}la Suki.
\newblock {Temporal complexity in clinical manifestations of lung disease}.
\newblock \emph{Journal of Applied Physiology}, 110\penalty0 (6):\penalty0
  1723--1731, 2011.
\newblock ISSN 87507587.
\newblock \doi{10.1152/japplphysiol.01297.2010}.

\bibitem[Halford et~al.(2015)Halford, Shiau, Desrochers, Kolls, Dean, Waters,
  Azar, Haas, Kutluay, Martz, Sinha, Kern, Kelly, Sackellares, and
  LaRoche]{Halford2015}
J.~J. Halford, D.~Shiau, J.~A. Desrochers, B.~J. Kolls, B.~C. Dean, C.~G.
  Waters, N.~J. Azar, K.~F. Haas, E.~Kutluay, G.~U. Martz, S.~R. Sinha, R.~T.
  Kern, K.~M. Kelly, J.~C. Sackellares, and S.~M. LaRoche.
\newblock {Inter-rater agreement on identification of electrographic seizures
  and periodic discharges in ICU EEG recordings}.
\newblock \emph{Clinical Neurophysiology}, 126:\penalty0 1661--1669, 2015.
\newblock ISSN 18728952.
\newblock \doi{10.1016/j.clinph.2014.11.008}.

\bibitem[Howbert et~al.(2014)Howbert, Patterson, Stead, Brinkmann, Vasoli,
  Crepeau, Vite, Sturges, Ruedebusch, Mavoori, Leyde, Sheffield, Litt, and
  Worrell]{Howbert2014}
J.~Jeffry Howbert, Edward~E. Patterson, S.~Matt Stead, Ben Brinkmann, Vincent
  Vasoli, Daniel Crepeau, Charles~H. Vite, Beverly Sturges, Vanessa Ruedebusch,
  Jaideep Mavoori, Kent Leyde, W.~Douglas Sheffield, Brian Litt, and Gregory~A.
  Worrell.
\newblock {Forecasting seizures in dogs with naturally occurring epilepsy}.
\newblock \emph{PLoS ONE}, 9\penalty0 (1):\penalty0 e81920, 2014.
\newblock ISSN 19326203.
\newblock \doi{10.1371/journal.pone.0081920}.

\bibitem[Jim{\'{e}}nez-Jim{\'{e}}nez et~al.(2015)Jim{\'{e}}nez-Jim{\'{e}}nez,
  Nekkare, Flores, Chatzidimou, Bodi, Honavar, Mullatti, Elwes, Selway,
  Valent{\'{i}}n, and Alarc{\'{o}}n]{Jimenez-Jimenez2015}
Diego Jim{\'{e}}nez-Jim{\'{e}}nez, Ramesh Nekkare, Lorena Flores, Katerina
  Chatzidimou, Istvan Bodi, Mrinalini Honavar, Nandini Mullatti, Robert D~C
  Elwes, Richard~P. Selway, Antonio Valent{\'{i}}n, and Gonzalo Alarc{\'{o}}n.
\newblock {Prognostic value of intracranial seizure onset patterns for surgical
  outcome of the treatment of epilepsy}.
\newblock \emph{Clinical Neurophysiology}, 126:\penalty0 257--267, 2015.
\newblock ISSN 18728952.
\newblock \doi{10.1016/j.clinph.2014.06.005}.

\bibitem[Jirsa et~al.(2014)Jirsa, Stacey, Quilichini, Ivanov, and
  Bernard]{Jirsa2014}
Viktor~K. Jirsa, William~C. Stacey, Pascale~P. Quilichini, Anton~I. Ivanov, and
  Christophe Bernard.
\newblock {On the nature of seizure dynamics}.
\newblock \emph{Brain}, 137:\penalty0 2210--2230, 2014.
\newblock ISSN 14602156.
\newblock \doi{10.1093/brain/awu133}.

\bibitem[Karoly et~al.(2018)Karoly, Kuhlmann, Soudry, Grayden, Cook, and
  Freestone]{Karoly2018b}
Philippa~J Karoly, Levin Kuhlmann, Daniel Soudry, David~B Grayden, Mark~J Cook,
  and Dean~R Freestone.
\newblock {Seizure pathways: a model-based investigation}.
\newblock \emph{PLoS Computational Biology}, 14\penalty0 (10):\penalty0
  e1006403, 2018.
\newblock \doi{10.26188/5b6a999fa2316}.

\bibitem[Kaufmann et~al.(2020)Kaufmann, Seethaler, Lauseker, Fan, Vollmar,
  Noachtar, and R{\'{e}}mi]{Kaufmann2020}
Elisabeth Kaufmann, Magdalena Seethaler, Michael Lauseker, Min Fan, Christian
  Vollmar, Soheyl Noachtar, and Jan R{\'{e}}mi.
\newblock {Who seizes longest? Impact of clinical and demographic factors}.
\newblock \emph{Epilepsia}, 61:\penalty0 1376--1385, 2020.
\newblock ISSN 15281167.
\newblock \doi{10.1111/epi.16577}.

\bibitem[Kim et~al.(2011)Kim, Cho, Lee, Joo, Hong, Hong, and Seo]{Kim2011dur}
Daeyoung Kim, Jae-Wook Cho, Jihyun Lee, Eun~Yeon Joo, Seung~Chyul Hong,
  Seung~Bong Hong, and Dae-Won Seo.
\newblock {Seizure Duration Determined by Subdural Electrode Recordings in
  Adult Patients with Intractable Focal Epilepsy}.
\newblock \emph{Journal of Epilepsy Research}, 1\penalty0 (2):\penalty0 57--64,
  2011.
\newblock ISSN 2233-6249.
\newblock \doi{10.14581/jer.11011}.

\bibitem[Kim and Park(2011)]{Kim2011}
Jingu Kim and Haesun Park.
\newblock {Fast nonnegative matrix factorization: an active-set-like method and
  comparisons}.
\newblock \emph{SIAM Journal on Scientific Computing}, 33\penalty0
  (6):\penalty0 3261--3281, 2011.
\newblock ISSN 1538-2656.
\newblock URL \url{http://www.ncbi.nlm.nih.gov/pubmed/11868488}.

\bibitem[Kim et~al.(2014)Kim, He, and Park]{Kim2014}
Jingu Kim, Yunlong He, and Haesun Park.
\newblock {Algorithms for nonnegative matrix and tensor factorizations: a
  unified view based on block coordinate descent framework}.
\newblock \emph{Journal of Global Optimization}, 58:\penalty0 285--319, 2014.
\newblock ISSN 09255001.
\newblock \doi{10.1007/s10898-013-0035-4}.

\bibitem[King-Stephens et~al.(2015)King-Stephens, Mirro, Weber, Laxer, {Van
  Ness}, Salanova, Spencer, Heck, Goldman, Jobst, Shields, Bergey, Eisenschenk,
  Worrell, Rossi, Gross, Cole, Sperling, Nair, Gwinn, Park, Rutecki, Fountain,
  Wharen, Hirsch, Miller, Barkley, Edwards, Geller, Berg, Sadler, Sun, and
  Morrell]{King-Stephens2015}
David King-Stephens, Emily Mirro, Peter~B Weber, Kenneth~D Laxer, Paul~C {Van
  Ness}, Vicenta Salanova, David~C Spencer, Christianne~N Heck, Alica Goldman,
  Barbara Jobst, Donald~C Shields, Gregory~K Bergey, Stephan Eisenschenk,
  Gregory~A Worrell, Marvin~A Rossi, Robert~E Gross, Andrew~J Cole, Michael~R
  Sperling, Dileep~R Nair, Ryder~P. Gwinn, Yong~D Park, Paul~A Rutecki,
  Nathan~B Fountain, Robert~E Wharen, Lawrence~J Hirsch, Ian~O Miller,
  Gregory~L Barkley, Jonathan~C Edwards, Eric~B Geller, Michel~J Berg, Toni~L
  Sadler, Felice~T Sun, and Martha~J Morrell.
\newblock {Lateralization of mesial temporal lobe epilepsy with chronic
  ambulatory electrocorticography}.
\newblock \emph{Epilepsia}, 56\penalty0 (6):\penalty0 959--967, 2015.
\newblock \doi{10.1111/epi.13010}.

\bibitem[Kini et~al.(2016)Kini, Davis, and Wagenaar]{Kini2016}
Lohith~G Kini, Kathryn~A. Davis, and Joost~B. Wagenaar.
\newblock {Data integration: combined imaging and electrophysiology data in the
  cloud}.
\newblock \emph{Neuroimage}, 124:\penalty0 1175--1181, 2016.
\newblock \doi{10.1038/nm.2451.A}.

\bibitem[Kudlacek et~al.(2021)Kudlacek, Chvojka, Kumpost, Hermanovska, Posusta,
  Jefferys, Maturana, Novak, Cook, Otahal, Hlinka, and Jiruska]{Kudlacek2021}
Jan Kudlacek, Jan Chvojka, Vojtech Kumpost, Barbora Hermanovska, Antonin
  Posusta, John~G.R. Jefferys, Matias~I. Maturana, Ondrej Novak, Mark~J. Cook,
  Jakub Otahal, Jaroslav Hlinka, and Premysl Jiruska.
\newblock {Long-term seizure dynamics are determined by the nature of seizures
  and the mutual interactions between them}.
\newblock \emph{Neurobiology of Disease}, 154:\penalty0 105347, 2021.
\newblock ISSN 1095953X.
\newblock \doi{10.1016/j.nbd.2021.105347}.

\bibitem[Lagarde et~al.(2019)Lagarde, Buzori, Trebuchon, Carron, Scavarda,
  Milh, McGonigal, and Bartolomei]{Lagarde2019}
Stanislas Lagarde, Sinziana Buzori, Agn{\`{e}}s Trebuchon, Romain Carron,
  Didier Scavarda, Mathieu Milh, Aileen McGonigal, and Fabrice Bartolomei.
\newblock {The repertoire of seizure onset patterns in human focal epilepsies:
  Determinants and prognostic values}.
\newblock \emph{Epilepsia}, 60:\penalty0 85--95, 2019.
\newblock ISSN 15281167.
\newblock \doi{10.1111/epi.14604}.

\bibitem[Lee and Seung(1999)]{Lee1999}
D~D Lee and H~S Seung.
\newblock {Learning the parts of objects by non-negative matrix factorization.}
\newblock \emph{Nature}, 401:\penalty0 788--791, 1999.
\newblock ISSN 0028-0836.
\newblock \doi{10.1038/44565}.
\newblock URL \url{http://www.ncbi.nlm.nih.gov/pubmed/10548103}.

\bibitem[Mantel(1967)]{Mantel1967}
Nathan Mantel.
\newblock {The detection of disease clustering and a generalized regression
  approach}.
\newblock \emph{Cancer Research}, 27\penalty0 (1):\penalty0 209--220, 1967.
\newblock ISSN 0028-0836.
\newblock \doi{10.1038/240498a0}.

\bibitem[Martinet et~al.(2015)Martinet, Ahmed, Lepage, Cash, and
  Kramer]{Martinet2015}
L.~E. Martinet, O.~J. Ahmed, K.~Q. Lepage, S.~S. Cash, and M.~A. Kramer.
\newblock {Slow spatial recruitment of neocortex during secondarily generalized
  seizures and its relation to surgical outcome}.
\newblock \emph{Journal of Neuroscience}, 35\penalty0 (25):\penalty0
  9477--9490, 2015.
\newblock ISSN 0270-6474.
\newblock \doi{10.1523/JNEUROSCI.0049-15.2015}.
\newblock URL
  \url{http://www.jneurosci.org/cgi/doi/10.1523/JNEUROSCI.0049-15.2015}.

\bibitem[Meisel et~al.(2015)Meisel, Schulze-Bonhage, Freestone, Cook,
  Achermann, and Plenz]{Meisel2015}
Christian Meisel, Andreas Schulze-Bonhage, Dean Freestone, Mark~James Cook,
  Peter Achermann, and Dietmar Plenz.
\newblock {Intrinsic excitability measures track antiepileptic drug action and
  uncover increasing/decreasing excitability over the wake/sleep cycle}.
\newblock \emph{Proceedings of the National Academy of Sciences}, 112\penalty0
  (47):\penalty0 14694--14699, 2015.
\newblock ISSN 0027-8424.
\newblock \doi{10.1073/pnas.1513716112}.
\newblock URL
  \url{http://www.pnas.org/cgi/doi/10.1073/pnas.1513716112{\%}5Cnhttp://www.pnas.org/lookup/doi/10.1073/pnas.1513716112}.

\bibitem[Mitsis et~al.(2020)Mitsis, Anastasiadou, Christodoulakis,
  Papathanasiou, Papacostas, and Hadjipapas]{Mitsis2020}
Georgios~D. Mitsis, Maria~N. Anastasiadou, Manolis Christodoulakis,
  Eleftherios~S. Papathanasiou, Savvas~S. Papacostas, and Avgis Hadjipapas.
\newblock {Functional brain networks of patients with epilepsy exhibit
  pronounced multiscale periodicities, which correlate with seizure onset}.
\newblock \emph{Human Brain Mapping}, 41:\penalty0 2059--2076, 2020.
\newblock ISSN 10970193.
\newblock \doi{10.1002/hbm.24930}.

\bibitem[Nevado-Holgado et~al.(2012)Nevado-Holgado, Marten, Richardson, and
  Terry]{Nevado-Holgado2012}
Alejo~J. Nevado-Holgado, Frank Marten, Mark~P. Richardson, and John~R. Terry.
\newblock {Characterising the dynamics of EEG waveforms as the path through
  parameter space of a neural mass model: application to epilepsy seizure
  evolution}.
\newblock \emph{NeuroImage}, 59:\penalty0 2374--2392, 2012.
\newblock ISSN 10538119.
\newblock \doi{10.1016/j.neuroimage.2011.08.111}.
\newblock URL \url{http://dx.doi.org/10.1016/j.neuroimage.2011.08.111}.

\bibitem[Panagiotopoulou et~al.(2021)Panagiotopoulou, Papasavvas, Schroeder,
  Taylor, and Wang]{Panagiotopoulou2021}
Mariella Panagiotopoulou, Christoforos Papasavvas, Gabrielle~M Schroeder, Peter
  Taylor, and Yujiang Wang.
\newblock Fluctuations in eeg band power at subject-specific timescales over
  minutes to days are associated with changes in seizure dynamics.
\newblock \emph{arXiv}, q-bio.NC:\penalty0 2012.07105, 2021.

\bibitem[Proix et~al.(2018)Proix, Jirsa, Bartolomei, Guye, and
  Truccolo]{Proix2018}
Timoth{\'{e}}e Proix, Viktor~K. Jirsa, Fabrice Bartolomei, Maxime Guye, and
  Wilson Truccolo.
\newblock {Predicting the spatiotemporal diversity of seizure propagation and
  termination in human focal epilepsy}.
\newblock \emph{Nature Communications}, 9:\penalty0 1088, 2018.
\newblock ISSN 20411723.
\newblock \doi{10.1038/s41467-018-02973-y}.
\newblock URL \url{http://dx.doi.org/10.1038/s41467-018-02973-y}.

\bibitem[Rossi et~al.(1994)Rossi, Colicchio, and Scerrati]{Rossi1994}
G.~F. Rossi, G.~Colicchio, and M.~Scerrati.
\newblock {Resection surgery for partial epilepsy. Relation of surgical outcome
  with some aspects of the epileptogenic process and surgical approach}.
\newblock \emph{Acta Neurochirurgica}, 130:\penalty0 101--110, 1994.
\newblock ISSN 00016268.
\newblock \doi{10.1007/BF01405509}.

\bibitem[Saggio et~al.(2020)Saggio, Crisp, Scott, Karoly, Kuhlmann, Nakatani,
  Murai, D{\"{u}}mpelmann, Schulze-Bonhage, Ikeda, Cook, Gliske, Lin, Bernard,
  Jirsa, and Stacey]{Saggio2020}
Maria~Luisa Saggio, Dakota Crisp, Jared Scott, Phillippa~J Karoly, Levin
  Kuhlmann, Mitsuyoshi Nakatani, Tomohiko Murai, Matthias D{\"{u}}mpelmann,
  Andreas Schulze-Bonhage, Akio Ikeda, Mark Cook, Stephen~V Gliske, Jack Lin,
  Christophe Bernard, Viktor Jirsa, and William Stacey.
\newblock {A taxonomy of seizure dynamotypes}.
\newblock \emph{Elife}, 9:\penalty0 e55632, 2020.
\newblock \doi{https://doi.org/10.7554/eLife.55632}.
\newblock URL \url{https://doi.org/10.7554/eLife.55632}.

\bibitem[Sakoe and Seibi(1978)]{Sakoe1978}
Hiroaki Sakoe and Chiba Seibi.
\newblock {Dynamic programming algorithm optimization for spoken word
  recognition}.
\newblock \emph{IEEE Transactions on Accoustics, Speech, and Signal
  Processing}, ASSP-26\penalty0 (1):\penalty0 43--49, 1978.

\bibitem[Salami et~al.(2020)Salami, Peled, Nadalin, Martinet, Kramer, Lee, and
  Cash]{Salami2020}
Pariya Salami, Noam Peled, Jessica~K. Nadalin, Louis~Emmanuel Martinet, Mark~A.
  Kramer, Jong~W. Lee, and Sydney~S. Cash.
\newblock {Seizure onset location shapes dynamics of initiation}.
\newblock \emph{Clinical Neurophysiology}, 131:\penalty0 1782--1797, 2020.
\newblock ISSN 18728952.
\newblock \doi{10.1016/j.clinph.2020.04.168}.

\bibitem[Salami et~al.(2021)Salami, Borzello, Kramer, Westover, and
  Cash]{Salami2021}
Pariya Salami, Mia Borzello, Mark~A Kramer, M~Brandon Westover, and Sydney~S
  Cash.
\newblock {Quantification of seizure termination patterns reveals limited
  pathways to seizure end}.
\newblock \emph{medRxiv}, page doi: 10.1101/2021.03.03.21252789, 2021.
\newblock URL
  \url{http://medrxiv.org/content/early/2021/03/05/2021.03.03.21252789.abstract}.

\bibitem[Sammon(1969)]{Sammon1969}
John~W. Sammon.
\newblock {A nonlinear mapping for data structure analysis}.
\newblock \emph{IEEE Transactions on Computers}, C-18\penalty0 (5):\penalty0
  401--409, 1969.
\newblock URL \url{http://repositorio.unan.edu.ni/2986/1/5624.pdf}.

\bibitem[Schindler et~al.(2007)Schindler, Leung, Elger, and
  Lehnertz]{Schindler2007a}
Kaspar Schindler, Howan Leung, Christian~E. Elger, and Klaus Lehnertz.
\newblock {Assessing seizure dynamics by analysing the correlation structure of
  multichannel intracranial EEG}.
\newblock \emph{Brain}, 130:\penalty0 65--77, 2007.
\newblock ISSN 00068950.
\newblock \doi{10.1093/brain/awl304}.

\bibitem[Schmeiser et~al.(2017)Schmeiser, Zentner, Steinhoff, Brandt,
  Schulze-Bonhage, Kogias, and Hammen]{Schmeiser2017}
B.~Schmeiser, J.~Zentner, B.~J. Steinhoff, A.~Brandt, A.~Schulze-Bonhage,
  E.~Kogias, and T.~Hammen.
\newblock {The role of presurgical EEG parameters and of reoperation for
  seizure outcome in temporal lobe epilepsy}.
\newblock \emph{Seizure}, 51:\penalty0 174--179, 2017.
\newblock ISSN 15322688.
\newblock \doi{10.1016/j.seizure.2017.08.015}.
\newblock URL \url{https://doi.org/10.1016/j.seizure.2017.08.015}.

\bibitem[Schroeder et~al.(2020)Schroeder, Diehl, Chowdhury, Duncan, de~Tisi,
  Trevelyan, Forsyth, Jackson, Taylor, and Wang]{Schroeder2020}
Gabrielle~M. Schroeder, Beate Diehl, Fahmida~A. Chowdhury, John~S. Duncan, Jane
  de~Tisi, Andrew~J. Trevelyan, Rob Forsyth, Andrew Jackson, Peter~N. Taylor,
  and Yujiang Wang.
\newblock Seizure pathways change on circadian and slower timescales in
  individual patients with focal epilepsy.
\newblock \emph{Proceedings of the National Academy of Sciences}, 117\penalty0
  (20):\penalty0 11048--11058, 2020.
\newblock ISSN 0027-8424.
\newblock \doi{10.1073/pnas.1922084117}.
\newblock URL \url{https://www.pnas.org/content/117/20/11048}.

\bibitem[Spencer et~al.(1981)Spencer, Spencer, Williamson, and
  Mattson]{Spencer1981}
Susan~S Spencer, Dennis~D Spencer, Peter~D Williamson, and Richard~H Mattson.
\newblock {Ictal effects of anticonvulsant medication withdrawal in epileptic
  patients}.
\newblock \emph{Epilepsia}, 22:\penalty0 297--307, 1981.

\bibitem[Tibshirani et~al.(2001)Tibshirani, Walther, and
  Hastie]{Tibshirani2001}
R~Tibshirani, G~Walther, and T~Hastie.
\newblock {Estimating the number of clusters in a data set via the gap
  statistic}.
\newblock \emph{Journal of the Royal Statistical Society: Series B (Statistical
  Methodology)}, 63:\penalty0 411--423, 2001.
\newblock ISSN 1369-7412.
\newblock \doi{10.1111/1467-9868.00293}.
\newblock URL
  \url{http://onlinelibrary.wiley.com/doi/10.1111/1467-9868.00293/abstract}.

\bibitem[Trevelyan et~al.(2007)Trevelyan, Sussillo, and Yuste]{Trevelyan2007}
Andrew~J Trevelyan, David Sussillo, and Rafael Yuste.
\newblock {Feedforward inhibition contributes to the control of epileptiform
  propagation speed.}
\newblock \emph{The Journal of Neuroscience}, 27\penalty0 (13):\penalty0
  3383--3387, 2007.
\newblock ISSN 0270-6474.
\newblock \doi{10.1523/JNEUROSCI.0145-07.2007}.

\bibitem[Truccolo et~al.(2011)Truccolo, Donoghue, Hochberg, Eskandar, Madsen,
  Anderson, Brown, Halgren, and Cash]{Truccolo2011}
Wilson Truccolo, Jacob~a Donoghue, Leigh~R Hochberg, Emad~N Eskandar, Joseph~R
  Madsen, William~S Anderson, Emery~N Brown, Eric Halgren, and Sydney~S Cash.
\newblock {Single-neuron dynamics in human focal epilepsy.}
\newblock \emph{Nature Neuroscience}, 14\penalty0 (5):\penalty0 635--641, 2011.
\newblock ISSN 1097-6256.
\newblock \doi{10.1038/nn.2782}.
\newblock URL \url{http://dx.doi.org/10.1038/nn.2782}.

\bibitem[Wagenaar et~al.(2013)Wagenaar, Brinkmann, Ives, Worrell, and
  Litt]{Wagenaar2013}
Joost~B. Wagenaar, Benjamin~H. Brinkmann, Zachary Ives, Gregory~A. Worrell, and
  Brian Litt.
\newblock {A multimodal platform for cloud-based collaborative research}.
\newblock \emph{International IEEE/EMBS Conference on Neural Engineering},
  pages 1386--1389, 2013.
\newblock ISSN 19483546.
\newblock \doi{10.1109/NER.2013.6696201}.

\bibitem[Wagner et~al.(2015)Wagner, Eskandar, Cosgrove, Madsen, Blum, Potter,
  Hochberg, Cash, and Truccolo]{Wagner2015}
Fabien~B. Wagner, Emad~N. Eskandar, G.~Rees Cosgrove, Joseph~R. Madsen,
  Andrew~S. Blum, N.~Stevenson Potter, Leigh~R. Hochberg, Sydney~S. Cash, and
  Wilson Truccolo.
\newblock {Microscale spatiotemporal dynamics during neocortical propagation of
  human focal seizures}.
\newblock \emph{NeuroImage}, 122:\penalty0 114--130, 2015.
\newblock ISSN 10959572.
\newblock \doi{10.1016/j.neuroimage.2015.08.019}.
\newblock URL \url{http://dx.doi.org/10.1016/j.neuroimage.2015.08.019}.

\bibitem[Wang et~al.(2017)Wang, Trevelyan, Valentin, Alarcon, Taylor, and
  Kaiser]{Wang2017}
Yujiang Wang, Andrew~J Trevelyan, Antonio Valentin, Gonzalo Alarcon, Peter~N
  Taylor, and Marcus Kaiser.
\newblock {Mechanisms underlying different onset patterns of focal seizures}.
\newblock \emph{PLoS Computational Biology}, 13\penalty0 (5):\penalty0
  e1005475, 2017.
\newblock ISSN 1553-7358.
\newblock \doi{10.1371/journal.pcbi.1005475}.
\newblock URL \url{http://dx.doi.org/10.1371/journal.pcbi.1005475}.

\bibitem[Wendling et~al.(2002)Wendling, Bartolomei, Bellanger, and
  Chauvel]{Wendling2002}
F.~Wendling, F.~Bartolomei, J.~J. Bellanger, and P.~Chauvel.
\newblock Epileptic fast activity can be explained by a model of impaired
  gabaergic dendritic inhibition.
\newblock \emph{European Journal of Neuroscience}, 15\penalty0 (9):\penalty0
  1499--1508, 2002.
\newblock \doi{https://doi.org/10.1046/j.1460-9568.2002.01985.x}.
\newblock URL
  \url{https://onlinelibrary.wiley.com/doi/abs/10.1046/j.1460-9568.2002.01985.x}.

\bibitem[Wenzel et~al.(2017)Wenzel, Hamm, Peterka, and Yuste]{Wenzel2017}
Michael Wenzel, Jordan~P. Hamm, Darcy~S. Peterka, and Rafael Yuste.
\newblock {Reliable and elastic propagation of cortical seizures in vivo}.
\newblock \emph{Cell Reports}, 19:\penalty0 2681--2693, 2017.
\newblock ISSN 22111247.
\newblock \doi{10.1016/j.celrep.2017.05.090}.
\newblock URL
  \url{http://linkinghub.elsevier.com/retrieve/pii/S2211124717307799}.

\bibitem[Wu et~al.(2016)Wu, Joseph, Hammonds, Celniker, Yu, and Frise]{Wu2016}
Siqi Wu, Antony Joseph, Ann~S. Hammonds, Susan~E. Celniker, Bin Yu, and Erwin
  Frise.
\newblock {Stability-driven nonnegative matrix factorization to interpret
  spatial gene expression and build local gene networks}.
\newblock \emph{Proceedings of the National Academy of Sciences}, 113\penalty0
  (16):\penalty0 4290--4295, 2016.
\newblock ISSN 0027-8424.
\newblock \doi{10.1073/pnas.1521171113}.
\newblock URL \url{http://www.pnas.org/lookup/doi/10.1073/pnas.1521171113}.

\end{thebibliography}

\newpage

\section*{Supplementary}
\renewcommand{\thefigure}{S\arabic{figure}}
\setcounter{figure}{0}
\counterwithin{figure}{subsection}
\counterwithin{table}{subsection}
\renewcommand\thesubsection{Supplementary S\arabic{subsection}}
\setcounter{subsection}{0}

\subsection{Subject metadata\label{suppl:subject_meta}}

Table \ref{emu_metadata} provides the following metadata for the \ac{EMU} patients:
\begin{itemize}
    \item \textbf{Hospital:} hospital at which the patient underwent presurgical monitoring.
    \item \textbf{Age:} age, in years, at the time of the presurgical monitoring.
    \item \textbf{Sex:} patient sex.
    \item \textbf{Hemisphere:} purported hemisphere of onset of the patient’s seizures, based on clinical findings.
    \item \textbf{Lobe: }purported lobe of onset of the patient’s seizures, based on clinical findings. Note that some patients had seizures arising from multiple lobes/at the boundary of two lobes (e.g., OP = occipital/parietal onset).
    \item \textbf{Pathology:} postoperative tissue pathology findings.  
    \item \textbf{ILAE surgical outcome:} patient surgical outcome according to the International League Against Epilepsy classification (1 = seizure free, 2 = only auras, 3+ = not seizure free). A dash indicates that the patient did not undergo surgery or their surgical outcome is unavailable. For IEEG Portal patients (MC and HUP hospitals), the surgical outcome provided by the database is given. For \ac{UCLH} patients, the 12 months post-surgical outcome is provided. 
    \item \textbf{Total recording time:} total duration of the presurgical intracranial recording time. 
    \item \textbf{\# seizures analysed:} number of the patient’s seizures analysed in this work.
    \item \textbf{\# electrodes analysed:} number of recording electrodes included in the analysis, after removing noisy electrodes.
    \item \textbf{Sampling frequencies:} sampling frequencies at which intracranial data was acquired and stored.
    \item \textbf{AED reduction performed:} whether patient \ac{AED}s were systematically reduced during the presurgical recording. A dash indicates that this information is unavailable.
\end{itemize}

Table \ref{nv_metadata} provides the following metadata for the NeuroVista patients:
\begin{itemize}
    \item \textbf{Age (yrs):} patient age in years.
    \item \textbf{Sex:} patient sex.
    \item \textbf{Age at diagnosis (yrs):} patient age when they were diagnosed with epilepsy, in years.
    \item \textbf{Lobe: }purported lobe of onset of the patient’s seizures, based on clinical findings. Note that some patients had seizures arising from multiple lobes (e.g., OP = occipital/parietal onset).
    \item \textbf{Previous resection}: whether the patient had undergone surgical resection prior to the chronic recording.
    \item \textbf{\# seizures analysed:} number of the patient’s seizures analysed in this work.
    \item \textbf{\# electrodes analysed:} number of recording electrodes included in the analysis after removing noisy electrodes.
    \item \textbf{Total recording time (days):} total duration of the intracranial recording time, in days. 
    \item \textbf{Sampling frequency:} sampling frequency at which intracranial data was acquired and stored.
\end{itemize}

Table \ref{canine_metadata} provides the following metadata for the dogs:
\begin{itemize}
    \item \textbf{Total recording time:} total duration of the \ac{iEEG} recording. 
    \item \textbf{\# seizures analysed:} number of the subjects’s seizures analysed in this work.
    \item \textbf{\# electrodes analysed:} number of recording electrodes included in the analysis, after removing noisy electrodes.
    \item \textbf{Sampling frequencies:} sampling frequencies at which intracranial data was acquired and stored.
\end{itemize}

\begin{table}
	    \centering
		\includegraphics[width=\textwidth]{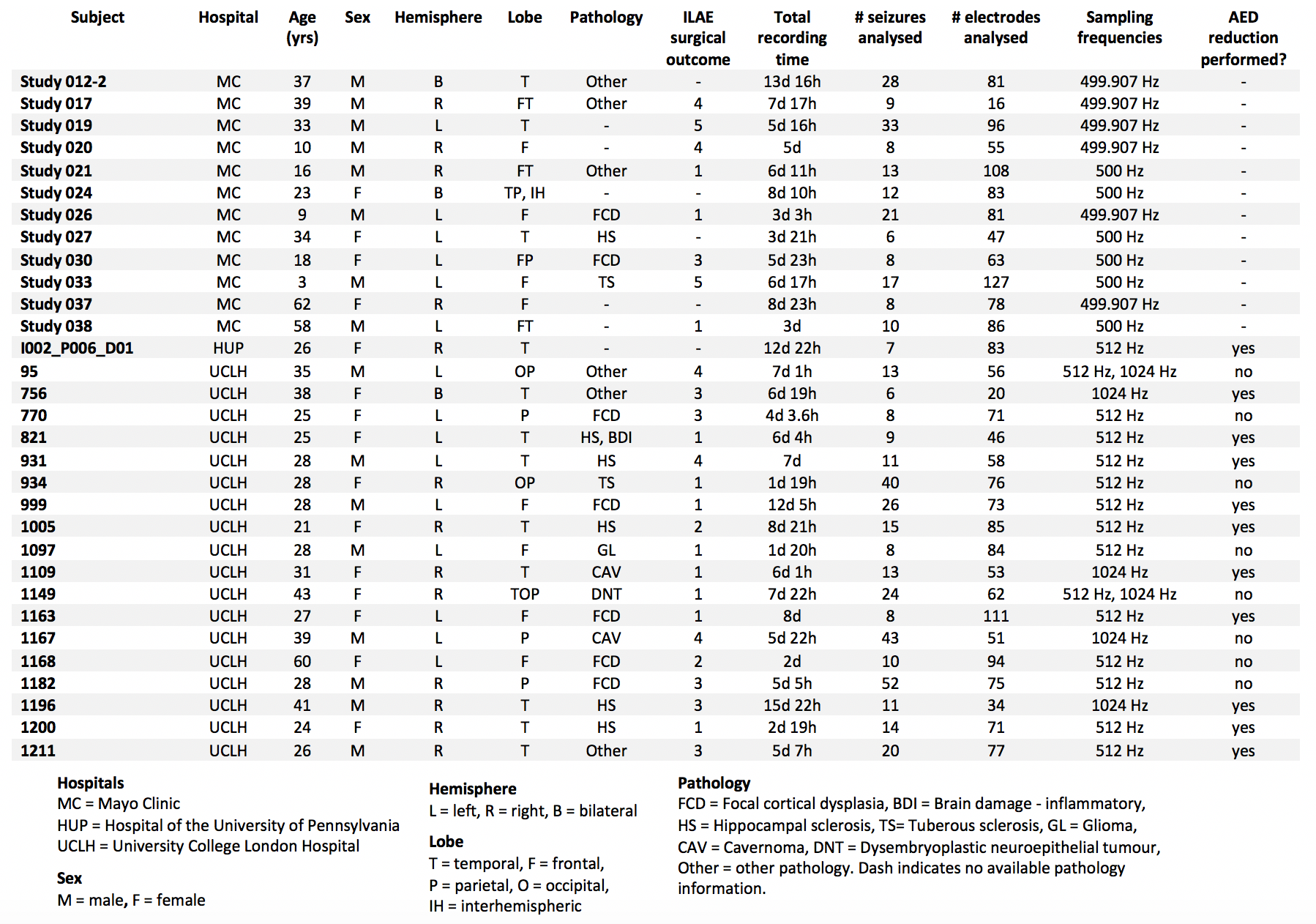}
		\caption[Metadata of \ac{EMU} patients]{\textbf{Metadata of \ac{EMU} patients.} Patient identifiers are listed under ``Subject.” IEEG Portal patients (MC and HUP hospitals) have the same identifier as the one used by the database. Metadata was extracted from the reports provided on the IEEG Portal (MC and HUP patients) or the patient clinical reports (UCLH patients). 
	}
		\label{emu_metadata}
\end{table}

\begin{table}[h]
	\centering
	\includegraphics[width=\textwidth]{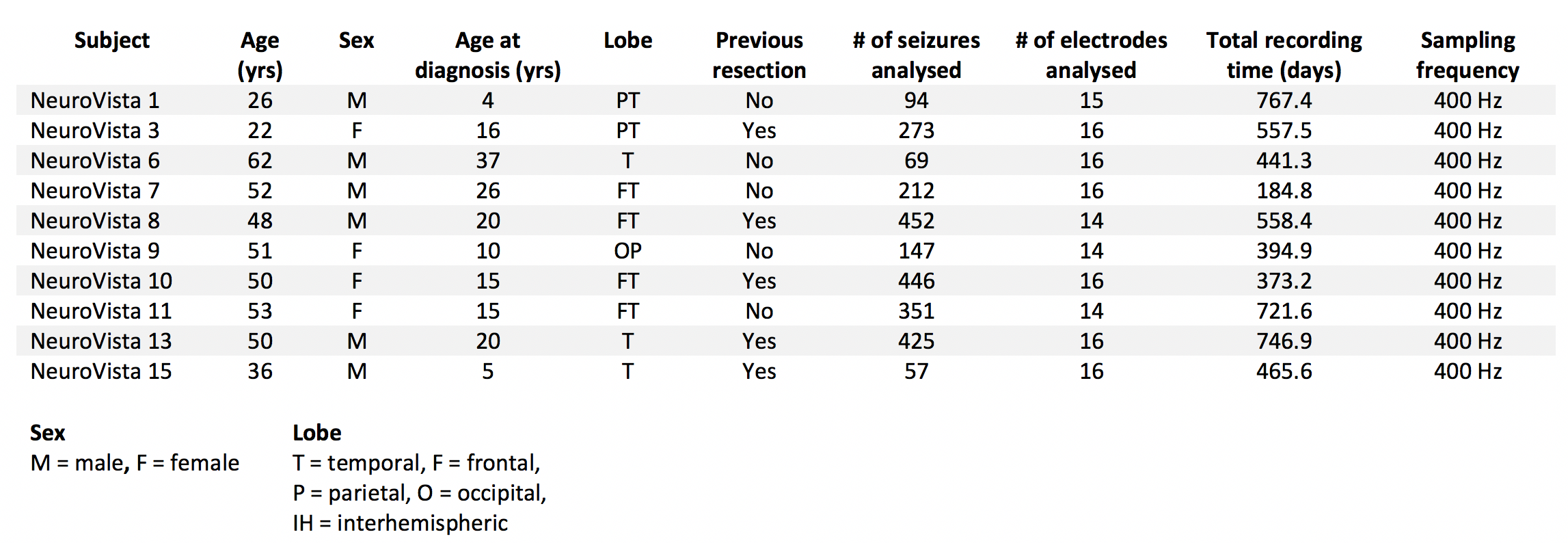}
	\caption[Metadata of NeuroVista patients]{\textbf{Metadata of NeuroVista patients.} Clinical metadata and patient demographics are reproduced from \cite{Cook2013}.  
	}
	\label{nv_metadata}
\end{table}

\begin{table} [h]
	\centering
	\includegraphics[width=\textwidth]{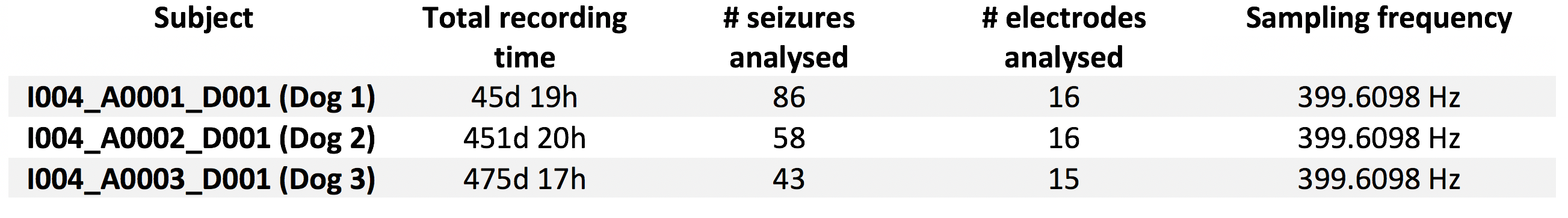}
	\caption[Metadata of dogs]{\textbf{Metadata of dogs.}
	}
	\label{canine_metadata}
\end{table}

\newpage 

\subsection{Dogs: identifying seizure terminations} \label{SS:canine_termination}

For recordings from the dogs, seizure onset was determined using the seizure onsets provided on the IEEG Portal; however, seizure termination times were not marked in this dataset. Therefore, seizure termination was identified algorithmically using an approach similar to \cite{Schindler2007a}. For each dog, the time periods around each marked seizure onset were extracted, beginning with 300s before seizure onset and ending with sufficient time after seizure onset to capture all seizure terminations, based on visual inspection (460s for Dog 1, 250s for Dog 2, and 150s for Dog 3). Because identifying seizure termination relied on reference preictal data, dog seizures were only included in the analysis if 1) there was at least 300s between the seizure start and the termination of the previous seizure, and 2) if the preictal period, defined as three minutes to one minute before seizure start, lacked large noisy or missing segments. 
    
After the preprocessing steps described in \ref{SS:preprocessing}, we identified the time period containing seizure activity for each channel in each \ac{iEEG} segment. Seizure activity was identified based on an increase in signal absolute slope, $S(t)$, compared to each seizure’s preictal period. The absolute slope $S$ of each channel $i$ was given by	

\begin{center}
$S_i(t) = | \frac{x_i(t) - x_i(t-1)}{ \Delta t}|$
\end{center}

where $x_i$ is the time series voltage value of channel $i$ and $\Delta t$ is size of the time step between successive \ac{iEEG} time points. $S_i(t)$ was then normalised to $S'_i(t)$ by dividing each time point by $\sigma_{i,pre}$, the standard deviation of the absolute slope of channel $i$ during the seizure’s preictal period, and smoothed by applying a 5s moving average sliding window. Channel $i$ was considered epileptic at time point $t$ if $S'_i (t)$ was greater than 2.5. Seizure termination was marked as the first time, following the clinically marked seizure start, when the number of epileptic channels fell below and remained below two channels for at least 1.5s.

\subsection{Significance test results for pathway dissimilarities/duration differences correlations} \label{S6:rho_sig}

\begin{figure}[h!]
	\centering
	\includegraphics[width=\textwidth]{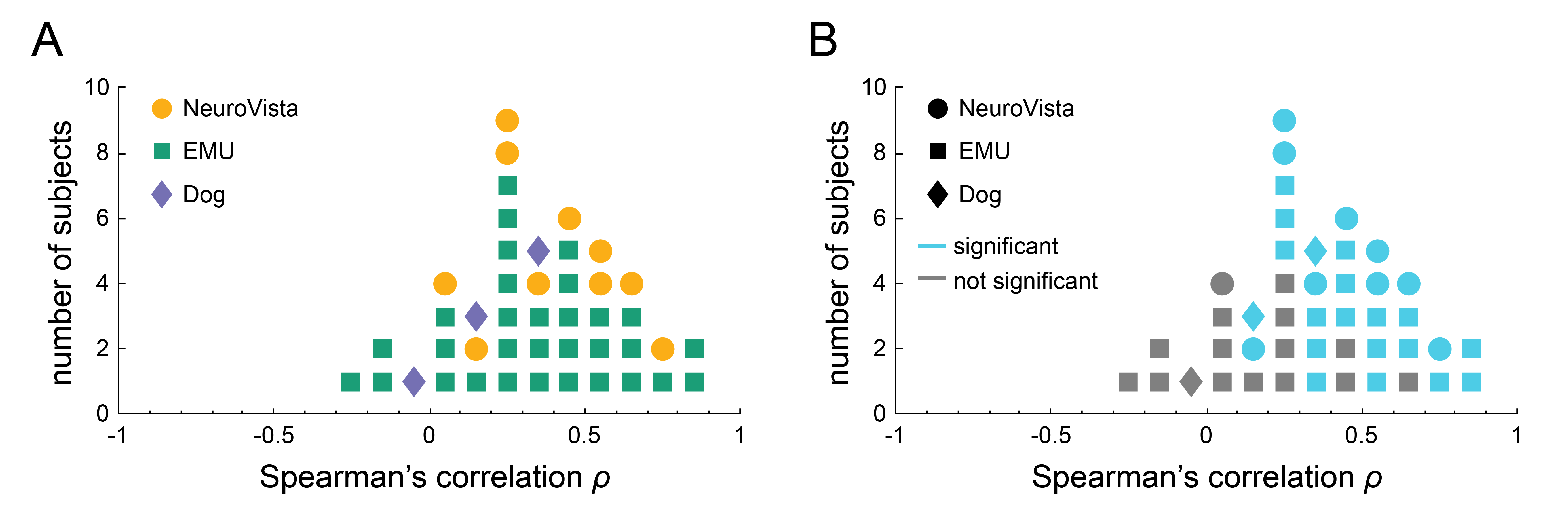}
	\caption[Significance test results for pathway dissimilarities/duration differences correlations]{\textbf{Significance test results for pathway dissimilarities/duration differences correlations.} A) Reproduction of Fig. \ref{F:rho}J showing the distribution of correlations between pathway dissimilarities and duration differences across subjects. Each marker corresponds to a subject, with the marker shape and colour indicating the subject's cohort. B) The same distribution as in A, but with markers now coloured by whether the subjects correlation was significant after \ac{FDR} correction for multiple comparisons.}
	\label{SF:rho_sig}
\end{figure}

Fig. \ref{SF:rho_sig} shows which correlations between pathway dissimilarites and duration differences were significant after \ac{FDR} correction for multiple comparisons. The Mantel test was used to test the statistical significance of each correlation (see Methods, section \ref{M7:fdr}). Note that the significance of each correlation depended not only on the strength of the correlation, but also the number of the seizures in each subject and the relationships between the subject's seizure pathways. As such, some relatively high correlations may not be significant and vice versa.  

\subsection{The relationship between pathway and duration variability does not depend on the range in either feature.} \label{S6:rho_vs_max}

\begin{figure}[h!]
	\centering
	\includegraphics[width=\textwidth]{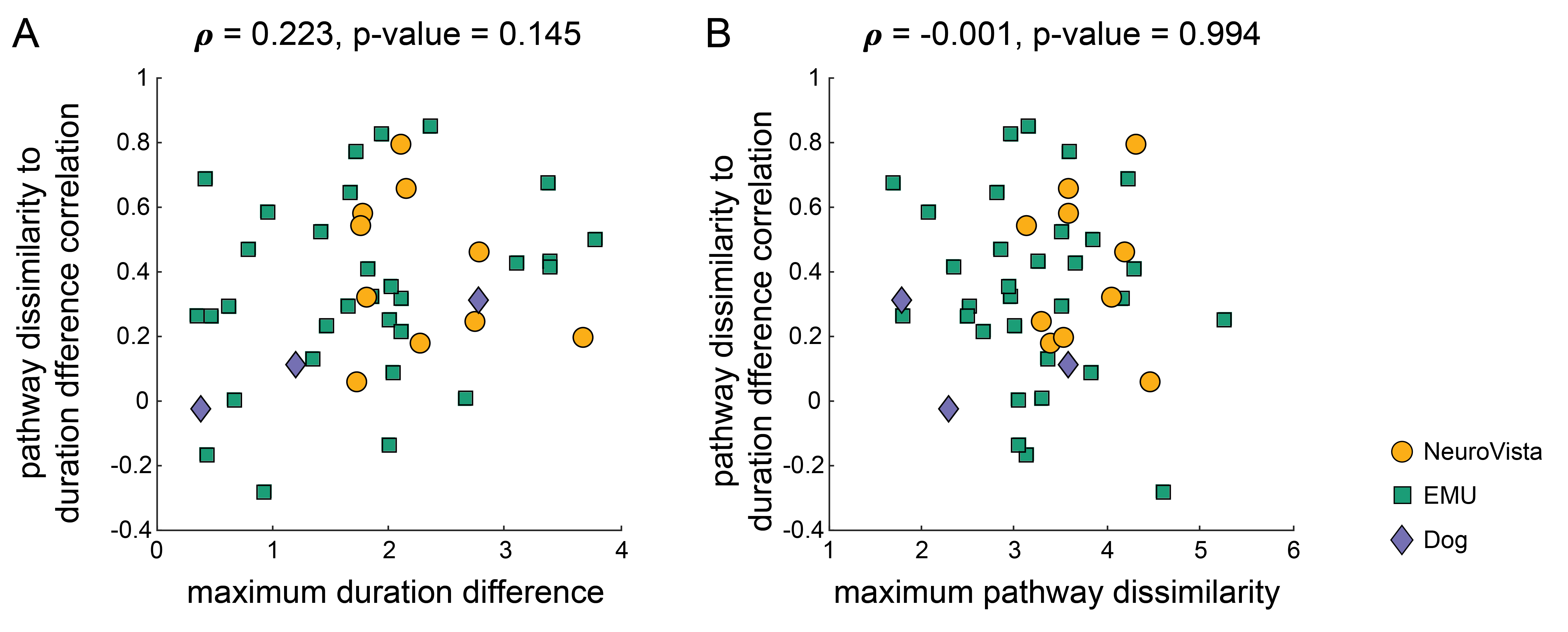}
	\caption[The relationship between pathway and duration variability does not depend on the range in either feature]{\textbf{The relationship between pathway and duration variability does not depend on the range in either feature.} A) The correlation between pathway dissimilarities and duration differences plotted versus the maximum duration difference of each subject. The colour and shape of each marker corresponds to the subject's cohort. B) The correlation between pathway dissimilarities and duration differences plotted versus the maximum pathway dissimilarity of each subject.}
	\label{SF:max_vs_rho}
\end{figure}

It is possible that the relationship between seizure pathways and durations requires the existence of a certain amount of variability in pathways and duration. For example, if seizure durations are highly consistent, then we would not expect there to be coinciding changes in seizure pathways. To test this hypothesis, we compared the correlation between pathway dissimilarities and duration differences to the maximum duration difference (Fig. \ref{SF:max_vs_rho}A) and maximum pathway dissimilarity (Fig. \ref{SF:max_vs_rho}B) in each subject. The maximum duration difference describes the greatest proportional change in seizure duration within each subject, while the maximum pathway dissimilarity captures the largest level of variability in the subject's seizure pathways. In other words, these measures serves as ranges for each subject's seizure durations and seizure pathways. 

In both cases, we found no significant relationship between the extent of variability and the correlation between pathway dissimilarities and duration differences. There was a weak, but insignificant, positive association between pathway dissimilarity/duration difference correlations and maximum durations differences. As such, there was a slight tendency for subjects with greater duration variability to have a stronger relationship between pathways and durations. Future work could investigate this relationship in a larger cohort. However, overall, the relationship between pathways and durations does not depend on the range of these features.

\subsection{Prevalence and features of elastic pathways and duplicate durations} \label{S6:elastic_duplications}

\begin{figure}[h!]
	\centering
	\includegraphics[width=\textwidth]{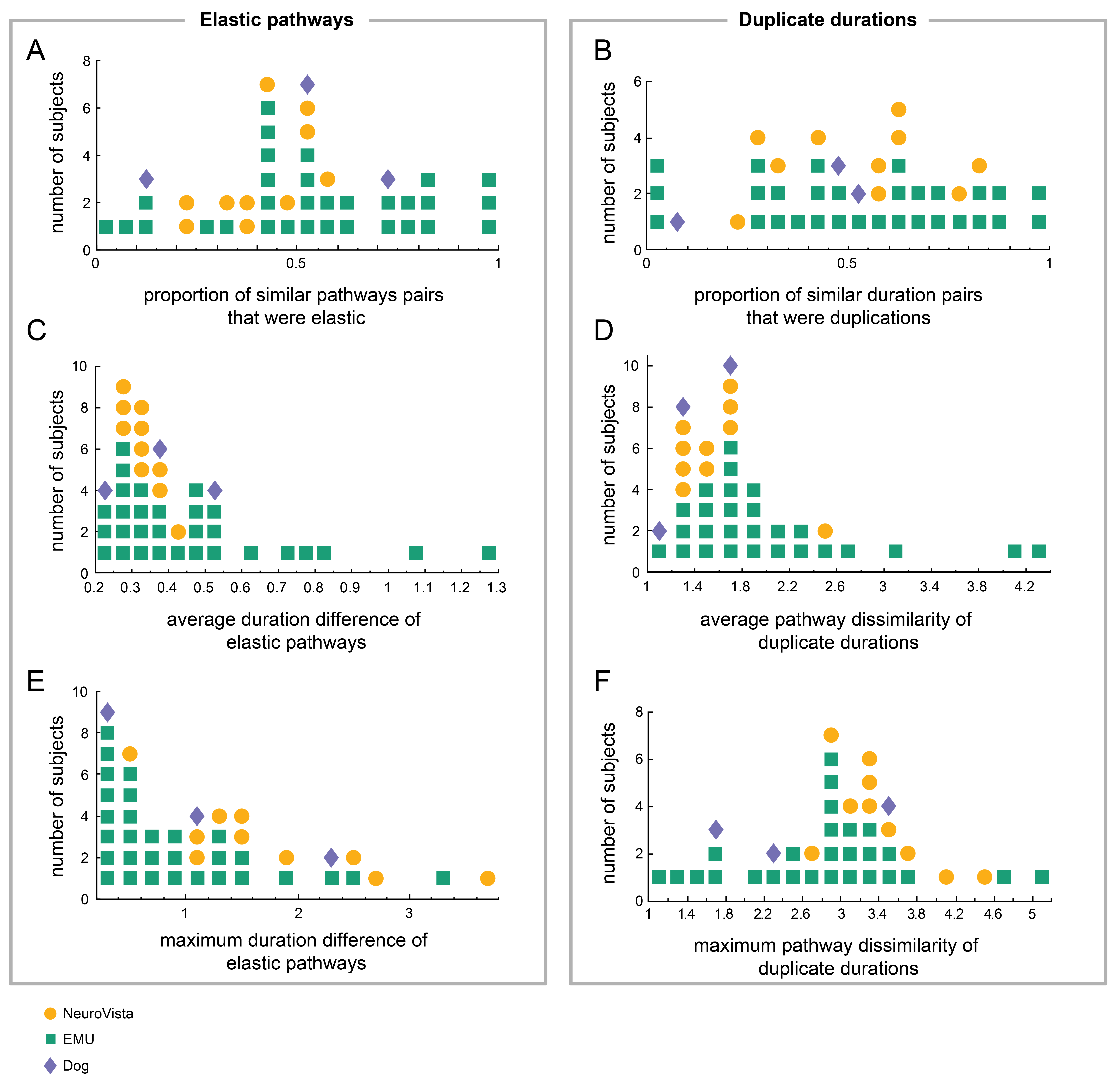}
	\caption[Prevalence and features of elastic pathways and duplicate durations]{\textbf{Prevalence and features of elastic pathways and duplicate durations.} In all plots, markers correspond to subjects and their colour and shape indicates the subjects' cohorts. A) Proportion of seizure pairs with similar pathways that were elastic (i.e., that had different durations) in each subject. B) Proportion of seizure pairs with similar durations that were duplications (i.e., that had different pathways). C,E) In subjects with elastic pathway pairs, the median (C) and maximum (E) duration difference of the elastic pathways. D,F) In subjects with duplicate duration pairs, the median (D) and maximum (F) pathway dissimilarity of the duplicate durations.}
	\label{SF:elastic_duplication}
\end{figure}

Fig. \ref{SF:elastic_duplication} describes the prevalence of elastic pathways and duplicate durations in each subject, as well as the level of elasticity and duplication (see Methods, section \ref{M7:elas}, for how elastic pathways and duplicate durations were defined). Fig. \ref{SF:elastic_duplication}A demonstrates that it was common for a high proportion of seizures with similar pathways to have different durations (i.e., elastic pathways). Likewise, Fig. \ref{SF:elastic_duplication}B reveals that in many subjects, seizures with similar durations had different pathways (i.e., they were duplicate durations). 

The level of pathway elasticity, which can be quantified by the duration differences of the elastic seizure pairs, varied across subjects (Fig. \ref{SF:elastic_duplication}C,E). On average, duration differences of 0.2 to 0.55 (equivalent to a $e^{0.2}$ to $e^{0.55} = 1.22$ to $1.73$ fold change in seizure duration), were common, but extremes of $e^{1} = 2.72$ or higher were also observed in many subjects. Thus, seizures with similar pathways could have drastically different durations. Likewise, the level of pathway dissimilarity between seizure pairs with similar durations varied across subjects (Fig. \ref{SF:elastic_duplication}D,F). Average pathway dissimilarities of approximately 1.2 to 2 were common, but higher dissimilarities of 3 or higher were also observed in many subjects. Therefore, seizures with similar durations could have very different pathways. 

\subsection{Comparison of duration populations and pathway dissimilarities\label{S6:rand_sig}} 

\begin{figure}[h!]
		\centering
		\includegraphics[width=\textwidth]{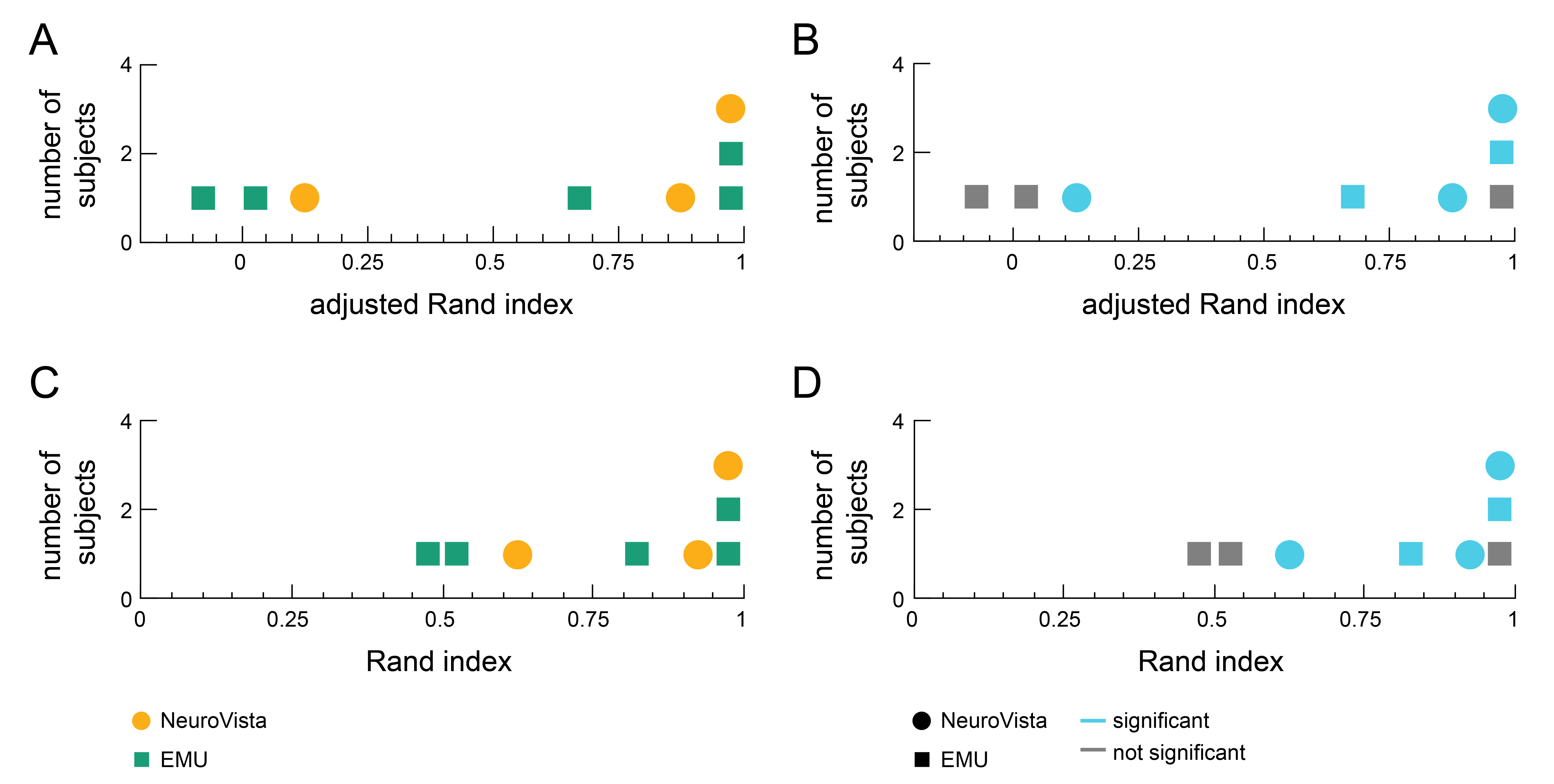}
		\caption[Comparison of duration populations and pathway dissimilarities]{\textbf{Comparison of duration populations and pathway dissimilarities.} A) Reproduction of Fig. \ref{F:dur_populations}I showing the distribution of adjusted Rand indices between duration populations and pathway clusters. Marker colour and shape correspond to each subject's cohort. B) Same as A, but marker colour indicates whether the adjusted Rand index of the subject was significant after correction for multiple comparisons. C-D) The corresponding Rand indices of the same subjects.}
		\label{SF:rand_sig}
\end{figure}

Fig. \ref{SF:rand_sig} shows the adjusted Rand indices (Fig. \ref{SF:rand_sig}A,B) and Rand indices (Fig. \ref{SF:rand_sig}C,D) between duration populations and pathway clusters in subjects with two duration populations (see Methods). The significance of each index for will depend on the index value as well as the number of seizures and cluster sizes. The Rand index is the proportion of seizure pairs that have the same relationship in both partitions (i.e., in the same cluster in both partitions or in different clusters in both partitions), and is therefore easily interpretable. However, the Rand Index greatly depends on the relative cluster sizes, making the adjusted Rand index a better measure for understanding the strength of the agreement of the two partitions. We therefore provide both measures here to evaluate the agreement between duration populations and pathway clusters.

\subsection{Clinical metadata comparisons\label{M7:clin}} 

To determine if patterns of pathway and duration variability were related, we compared nine subject-specific pathway and duration measures to four clinical variables in the \ac{EMU} patients. The pathway and duration measures were

\begin{enumerate}
	\item The correlation between the subject's pathway dissimilarities and duration differences.
	\item The subject's maximum duration difference.
	\item The subject's median duration difference.
	\item The subject's longest seizure.
	\item The subject's median seizure duration.
	\item The proportion of the subject's seizure pairs with similar pathways that were elastic.
	\item The proportion of the subject's seizure pairs with similar durations that were duplicate durations.
	\item The median duration difference of the subject's elastic pathways. 
	\item The median pathway dissimilarity of the subject's duplicate durations.
\end{enumerate} 

These measures were compared to
\begin{enumerate}
	\item The patients' \ac{ILAE} surgical outcomes ($n = 26$).
	\item The patients' disease durations ($n = 31$).
	\item Whether the patient had temporal ($n = 12$) or extratemporal ($n = 15$) epilepsy.
	\item Whether the patient had seizures with left ($n = 15$) or right ($n = 13$) hemisphere onset.
\end{enumerate}

\ac{ILAE} surgical outcomes and disease duration were associated with the variability measures using Spearman's correlation. The variability measures of temporal versus extratemporal patients and left versus right hemisphere onset patients were compared using Wilcoxon rank sum tests. 

There were no significant relationships between seizure pathway/duration measures and surgical outcome after \ac{FDR} correction for multiple comparisons, although there was a trend (Spearman's correlation $\rho = 0.45, p = 0.0209$) for surgical outcome to worsen as median seizure duration increased. This association may have been driven by other clinical factors that influence both surgical outcome and seizure duration, such as whether a patient has focal to bilateral tonic-clonic seizures \citep{Kaufmann2020,Dobesberger2015,Baud2015} and the localisation of the epileptogenic zone \citep{Kim2011dur,DeTisi2011}. There were also no significant differences between patients with different onset locations, and the only trend was for patients with temporal onset to have higher median seizure duration than patients with extratemporal onset  (Wilcoxon rank sum test, $p = 0.0381$), consistent with previous findings \citep{Kim2011dur}.

Disease duration was significantly inversely correlated to median seizure duration (Spearman's correlation $\rho = -0.56, p = 0.0011$), with median seizure duration decreasing as disease duration increased. Additionally, the tendency of similar seizure pathways to be elastic significantly decreased with disease duration (Spearman's correlation $\rho = -0.51, p = 0.0038$). Disease duration at the time of the \ac{iEEG} recording was not sampled at a random time for each patient, but determined by the clinical decision to undergo presurgical monitoring. Therefore, disease duration in our cohort could also be associated with clinical considerations such as seizure severity and the patient's level of antiepileptic medication resistance. As such factors could also influence measures such as seizure duration, it is difficult to interpret the observed associations with disease duration.

\end{document}